\documentclass[journal]{IEEEtran}
\usepackage{cite}  % Reference
\usepackage{graphicx}
\usepackage{algorithm}  % pseudo code
\usepackage{algpseudocode}  % pseudo code
\usepackage{latexsym}
\usepackage{amsmath,amssymb,amsfonts}  % math
\usepackage{bm}  % boldsymbol
\usepackage{xcolor}
\usepackage[acronym,shortcuts]{glossaries}
\usepackage[font=footnotesize]{subcaption}
\usepackage[font=footnotesize]{caption}

\newacronym{MIMO}{MIMO}{multiple-input-multiple-output}
\newacronym{SISO}{SISO}{single-input-single-output}
\newacronym{OFDM}{OFDM}{orthogonal frequency-division multiplexing}
\newacronym{ULA}{ULA}{uniform linear array}
\newacronym{UE}{UE}{user equipment}
\newacronym{BS}{BS}{base station}
\newacronym{CP}{CP}{cyclic prefix}
\newacronym{AWGN}{AWGN}{additive white Gaussian noise}
\newacronym{AoA}{AoA}{angles of arrival}
\newacronym{AoD}{AoD}{angles of departure}
\newacronym{LS}{LS}{least squares}
\newacronym{OMP}{OMP}{orthogonal matching pursuit}
\newacronym{SBL}{SBL}{sparse Bayesian learning}
\newacronym{RIP}{RIP}{restricted isometry property}
\newacronym{RIC}{RIC}{restricted isometry constant}
\newacronym{ISTA}{ISTA}{iterative shrinkage thresholding algorithm}
\newacronym{NMSE}{NMSE}{normalized mean-squared error}
\newacronym{LMMSE}{LMMSE}{linear minimum mean-squared error}
\newacronym{MMSE}{MMSE}{minimum mean-squared error}
\newacronym{BER}{BER}{bit error rate}
\newacronym{BLER}{BLER}{block error rate}
\newacronym{GA}{GA}{genetic algorithm}
\newacronym{SVD}{SVD}{singular value decomposition}
\newacronym{CDF}{CDF}{cumulative distribution function}
\newacronym{mmWave}{mmWave}{milimeter-wave}
\newacronym{cmWave}{cmWave}{centimeter-wave}
\newacronym{sub-THz}{sub-THz}{sub-terahertz}
\newacronym{SIDCO}{SIDCO}{successive iterative decorrelation by convex optimization}
\newacronym{QC-SIDCO}{QC-SIDCO}{quadratic complex-SIDCO}
\newacronym{CS}{CS}{compressed sensing}
\newacronym{MIP}{MIP}{mixed integer programming}
\newacronym{FLOPs}{FLOPs}{floating operations}
\newacronym{Adam}{Adam}{adaptive moment estimation}
\newacronym{SNR}{SNR}{signal-to-noise ratio}
\newacronym{CIR}{CIR}{channel impulse response}
\newacronym{Tx}{Tx}{transmitter}
\newacronym{Rx}{Rx}{receiver}
\newacronym{CDS}{CDS}{cyclic difference set}
\newacronym{QP}{QP}{quadratic programming}
\newacronym{NOMA}{NOMA}{non-orthogonal multiple access}
\newacronym{PRB}{PRB}{physical resource Block}
\newacronym{SE}{SE}{spectral efficiency}
\newacronym{GAMP}{GAMP}{generalized approximate message passing}
\newacronym{GAMP-SBL}{GAMP-SBL}{generalized approximate message passing-based sparse Bayesian learning}
\newacronym{RE}{RE}{resource element}
\newacronym{MINLP}{MINLP}{mixed‐integer nonlinear program}

\begin{document}
 
\title{Joint Pilot Allocation and Sequence Design for MIMO-OFDM Systems With Channel Sparsity}

\author{Kabuto Arai,~\IEEEmembership{Graduate Student Member, IEEE},  Koji Ishibashi,~\IEEEmembership{Senior Member, IEEE},\\ Hiroki Iimori,~\IEEEmembership{Member, IEEE}, Yuto Hama,~\IEEEmembership{Member, IEEE}, Paulo Valente Klaine, and Szabolcs Malomsoky
% <-this % stops a space
% 
\thanks{Kabuto Arai and Koji Ishibashi are with the Advanced Wireless and Communication Research Center (AWCC), The University of Electro-Communications, Tokyo 182-8285, Japan (e-mail: k.arai@awcc.uec.ac.jp, koji@ieee.org)}
\thanks{Hiroki Iimori, Yuto Hama, Paulo Valente Klaine, and Szabolcs Malomsoky are with Ericsson Research, Ericsson Japan K.K., Yokohama, 220-0012(e-mail: \{hiroki.iimori, yuto.hama, paulo.valente.klaine, szabolcs.malomsoky\}@ericsson.com)}
~\nocite{2024Arai_NearCE}
}

\markboth{Journal of \LaTeX\ Class Files,~Vol.~14, No.~8, August~2021}%
{Shell \MakeLowercase{\textit{et al.}}: Sample article using IEEEtran.cls for IEEE Journals}


\maketitle

\begin{abstract}  % (max: 200 words)
    This paper proposes a joint optimization of pilot subcarrier allocation and non-orthogonal sequence for \acf{MIMO}-\acf{OFDM} systems under \acf{CS}-based channel estimation exploiting delay and angle sparsity.
    Since the performance of CS-based approaches depends on a coherence metric of the sensing matrix in the measurement process, we formulate a joint optimization problem to minimize this coherence.
    Due to the discrete nature of subcarrier allocation, a straightforward formulation of the joint optimization results in a \acf{MINLP}, which is computationally intractable due to the combinatorial explosion of allocation candidates.
    To overcome the intractability of discrete variables, we introduce a block sparse penalty for pilots across all subcarriers, which ensures that the power of some unnecessary pilots approaches zero. 
    This framework enables joint optimization using only continuous variables.
    In addition, we propose an efficient computation method for the coherence metric by exploiting the structure of the sensing matrix, which allows its gradient to be derived in closed form, making the joint optimization problem solvable in an efficient way via a gradient descent approach.
    Numerical results confirm that the proposed pilot sequence exhibits superior coherence properties and enhances the CS-based channel estimation performance.
\end{abstract}

\glsresetall

\begin{IEEEkeywords}
    Channel estimation, compressed sensing, delay-angle sparsity, non-orthogonal sequence design, pilot design, subcarrier allocation.
\end{IEEEkeywords}

\IEEEpeerreviewmaketitle

\glsresetall

\section{Introduction}

To achieve high throughput and capacity in future wireless communication systems, the extensive spectral resources available in mid to high-frequency bands are crucial~\cite{2021Tataria_6G_syrvey, 2024Katwe_cmwave_subTHz}.
However, at higher frequencies, attenuation is even more significant due to limited diffraction, frequent blockages, and molecular absorption~\cite{2015Rappaport_mmWave_model, 2021Tarboush_TeraMIMO}. 
To counteract these issues, on the one hand, increasingly large antenna arrays are utilized at the \ac{BS} to generate narrow beams towards the users, in what is known as beamforming. 
On the other hand, increasing the number of antennas also increases the number of required pilots for channel estimation, reducing user throughput. 
As such, to reduce the pilot overhead, channel estimation approaches, leveraging channel sparsity in the delay-angle domain, have been proposed in~\cite{2017Venugopal_CE_delay_hybrid, 2018Mo_CE_ADC, 2018Wang_CS_CE, 2019_CE_delay_turbo, 2023Liu_CE_structured_MP, 2024Uchimura_Tracking}.
The channel sparsity is attributed to the fact that the channel is composed of a few dominant delay paths~\cite{2024Liu_model_CmMmST}.
These channel estimation methods rely on \ac{CS} techniques, which can significantly reduce pilot overhead compared to classical channel estimation
approaches based on \ac{LS} and \ac{LMMSE}.

Since classical and CS-based channel estimation methods are generally based on pilot signals, their estimation accuracy depends on the pilot subcarrier allocation and pilot sequence design.
Therefore, to improve channel estimation accuracy, many studies have investigated both pilot allocation and sequence design, albeit in a separate manner, for both classical and CS-based approaches.
% 
% Allocation (LS)
In~\cite{2003Barhumi_pilot_alloc_LS, 2003Hlaing_pilot_alloc_LS}, optimal subcarrier pilot allocation has been investigated for classical channel estimation methods based on \ac{LS} and \ac{LMMSE} in both \ac{SISO}-OFDM ~\cite{2006Zhang_pilot_alloc_LS, 2002Adireddy_pilot_alloc_LS} and MIMO-OFDM systems.
These studies demonstrated that a uniformly spaced pilot subcarrier allocation in the frequency domain is optimal in terms of \ac{MMSE} and capacity.
However, this allocation is not optimal for CS-based channel estimation.

% % Allocation (CS)
In CS-based estimation, the structure of the sensing matrix in the measurement equation significantly influences the performance of sparse signal recovery~\cite{2012Yonina_CS_book, 2011Duarte_CS_theory}. 
Since the sensing matrix depends on both the subcarrier allocation and the pilot sequence, it is crucial to carefully design these elements to enhance channel estimation accuracy.
One of the key properties of the sensing matrix is the \ac{RIP}~\cite{2012Yonina_CS_book, 2011Duarte_CS_theory}.
If a sensing matrix satisfies the RIP under certain conditions, it guarantees the exact recovery of a sparse signal.
However, because evaluating whether a given sensing matrix satisfies the RIP is computationally expensive~\cite{2012Yonina_CS_book, 2011Duarte_CS_theory}, it is impractical to design a sensing matrix solely based on RIP.
To design a sensing matrix with practical complexity, coherence metrics, such as mutual coherence and total coherence, are commonly used as alternatives to the RIP.
The design of sensing matrices based on coherence metrics has been extensively studied across various fields, including radar systems~\cite{2011Yu_SensingDesign_Rador, 2017Obermeier_SensingDesign_Eimag}, image processing~\cite{2022Jiang_SensingDesign_Imag, 2009Duarte_SensingDesign_Imag, 2011Zelnik_SensingDesign_Imag}, RIS-based communication systems~\cite{2023Chen_SensingDesign_RIS}, beamforming design~\cite{2019Stoica_BeamDesign, 2022Ge_BeamDesign}, and antenna design \cite{2023Bangun_SensingDesign_Ant}.

% CS-based pilot allocation (SISO)
In the context of OFDM systems, the optimization of subcarrier pilot allocation for sensing matrix design, based on coherence metrics, has been extensively studied in~\cite{2012Qi_allocation_SISO_OFDM,2012Pakrooh_allocation_SISO_OFDM, 2020Nie_allocation_SISO_OFDM, 2015Qi_allocation_SISO_OFDM, 2019Xiao_JMCTC}.
These studies focus on identifying the optimal subcarrier allocation in SISO-OFDM systems for CS-based channel estimation methods that leverage channel sparsity in the delay domain, where the \ac{CIR} length is constrained.
The optimal subcarrier allocation in CS-based estimation corresponds to a \ac{CDS}~\cite{2012Qi_allocation_SISO_OFDM, 2005Pengfei_CDS_Welch_bound}, which contrasts with the equally spaced subcarrier allocation used in classical estimation methods such as \ac{LS} and \ac{LMMSE}.
However, \ac{CDS}-based optimal allocation is limited to specific combinations of the total number of subcarriers and pilot subcarriers, restricting its general applicability.
Therefore, studies~\cite{2012Qi_allocation_SISO_OFDM,2012Pakrooh_allocation_SISO_OFDM, 2020Nie_allocation_SISO_OFDM, 2015Qi_allocation_SISO_OFDM, 2019Xiao_JMCTC} have proposed discrete optimization techniques to optimize subcarrier pilot allocation for any number of subcarriers.

% CS-based pilot allocation (MIMO)
The extension of pilot subcarrier allocation to MIMO-OFDM systems has been studied in~\cite{2015Qi_allocation_MIMO_OFDM, 2013He_allocation_MIMO_OFDM, 2016He_allocation_MIMO_OFDM, 2022Li_allocation_MIMO_OFDM, 2025Zhou_allocation_MIMO_OFDM}.
To prevent pilot contamination at the \ac{Rx} side, these studies assume orthogonal pilot allocation among \ac{Tx} antennas. Specifically, a \ac{Tx} antenna transmits a pilot signal on the $k$-th subcarrier,
while all the other \ac{Tx} antennas remain silent on that subcarrier. 
Although this orthogonal pilot transmission mitigates pilot contamination between \ac{Tx} antennas, it requires a large number of time-frequency resources to estimate channel coefficients for all \ac{Tx} antennas, especially in massive antenna systems.
To address the pilot overhead problem, the authors in~\cite{2017Mohammadian_allocation_MIMO_OFDM} proposed an optimized pilot subcarrier allocation for multiple \ac{Tx} antennas, by allowing non-orthogonal transmission.
However, since this method optimizes subcarrier allocation by exploiting only the delay-domain sparsity without considering the angle-domain sparsity, thus, the pilot overhead reduction is limited, especially as the number of Tx antennas increases.
Although it has been demonstrated that exploiting both delay and angle sparsity within CS frameworks can significantly reduce pilot overhead for channel estimation~\cite{2017Venugopal_CE_delay_hybrid, 2018Mo_CE_ADC, 2018Wang_CS_CE, 2019_CE_delay_turbo, 2023Liu_CE_structured_MP, 2024Uchimura_Tracking}, the aforementioned studies have not addressed pilot design specifically tailored for CS-based estimation frameworks that leverage the delay-angle sparsity.

% Sequence design
To enable efficient pilot transmission in MIMO-OFDM systems, it is essential to optimally design a non-orthogonal pilot sequence.
The studies~\cite{2016Rusu_SIDCO, 2018Rusu_CSIDCO, 2018Stoica_frame_VTC, 2019Stoica_frame_NOMA_Access, 2022Iimori_bilinear_Grant_free} proposed optimization methods for designing non-orthogonal sequences aimed at mitigating contamination between sequences.
In~\cite{2016Rusu_SIDCO, 2018Rusu_CSIDCO}, non-orthogonal sequences are designed by minimizing mutual coherence, ensuring that each sequence is as orthogonal as possible to the others, via an optimization technique referred to as \ac{SIDCO}.
The authors in~\cite{2018Stoica_frame_VTC, 2022Iimori_bilinear_Grant_free} extend SIDCO to the complex space, and its optimization problem is transformed into a \ac{QP} problem, which can be efficiently solved using generic solvers~\cite{2010Mattingley_RealTime_cvx}.
The non-orthogonal sequences designed using the \ac{QC-SIDCO} closely approach the lower bound of mutual coherence, commonly known as the Welch bound~\cite{1974Welch_bound}.
The effectiveness of the sequence designed by QC-SIDCO has been validated in terms of CS-based channel estimation across various fields, such as grant-free \ac{NOMA} systems~\cite{2019Stoica_frame_NOMA_Access, 2022Iimori_bilinear_Grant_free} and mmWave MIMO systems~\cite{2018Stoica_frame_VTC, 2019Stoica_frame_mmWave_Acess}.
However, these studies focus solely on the design of non-orthogonal sequences, without considering subcarrier pilot allocation.
Since the performance of CS-based channel estimation depends on both the design of non-orthogonal pilot sequences and the subcarrier allocation, a joint optimization of both the sequence design and allocation could potentially lead to better channel estimation performance.
To the best of our knowledge, no existing work has jointly addressed both subcarrier allocation and non-orthogonal sequence design under CS-based channel estimation, leveraging channel sparsity in the delay-angle domain.

% Contributions
This paper addresses joint pilot allocation and sequence design within the framework of CS-based channel estimation, in response to the aforementioned challenges.  
The main contributions of this article are summarized as follows:
\begin{itemize}
    \item \textbf{Optimal joint design of pilot subcarrier allocation and non-orthogonal pilot sequence}:
        To reduce the pilot overhead for channel estimation in MIMO-OFDM systems, we assume non-orthogonal pilot transmission and channel estimation exploiting the delay-angle sparsity.
        We formulate an optimization problem to jointly design subcarrier allocation and non-orthogonal sequences to minimize the coherence metric of the sensing matrix.
        The optimization problem is formulated as a \ac{MINLP} because subcarrier allocation requires discrete variables, while sequence design requires continuous variables.
        Solving the MINLP via brute-force search is computationally infeasible due to the combinatorial explosion in possible subcarrier allocations and the necessity of simultaneously designing non-orthogonal sequences.
        Thus, we set pilot sequences across all subcarriers rather than a limited number of subcarriers, and introduce a block sparse penalty to induce the power of unnecessary pilot subcarriers to approach zero.
        This framework enables the joint optimization using only continuous variables, without requiring integer variables.

    \item \textbf{Efficient computation of the coherence metric and its gradient for pilot design}:
        For efficient channel estimation via a CS-based method in the delay-angle domain, it is necessary to design the sensing matrix from dictionary matrices representing \ac{AoA}, \ac{AoD}, and delay, based on discrete grids that quantize the delay-angle domain.
        Because the sensing matrix has a three-dimensional grid structure encompassing AoA, AoD, and delay, its size grows substantially, resulting in increased computational complexity when minimizing the coherence metric.
        To address this issue, we decompose the coherence metric by leveraging the structure of the sensing matrix based on the properties of the Kronecker product.
        This decomposition enables efficient computation of the coherence metric independently of the dictionary matrix for AoD grids.
        Utilizing this decomposition, the gradient of the coherence metric can be derived in closed form, and the optimization problem can be efficiently solved using gradient descent.
\end{itemize}

% Organization of the paper
The rest of the paper is organized as follows.
Section~\ref{sec:system_model} describes the channel model with delay-angle sparsity and the pilot transmission scheme in a MIMO-OFDM system.
Section~\ref{sec:channel_est} presents the formulation of the sensing matrix and the channel estimation strategy, leveraging channel sparsity in the delay-angle domain.
Section~\ref{sec:prop} formulates the optimization problem for joint design of pilot allocation and pilot sequence.
Section~\ref{sec:simulation} provides numerical results. 
Finally, Section~\ref{sec:conclusion} concludes the paper.

% Notation
\textit{Notation}:
In this paper, the following notations are used. 
Bold lowercase letters denote vectors, and bold uppercase letters denote matrices.
The notation $(\cdot)^*$, $(\cdot)^\mathrm{T}$, and $(\cdot)^\mathrm{H}$ represent conjugate, transpose, and conjugate transpose, respectively.
$\mathbf{0}_{N \times M}$ and $\mathbf{I}_N$ denote the $N \times M$ zero matrix and the $N \times N$ identity matrix. 
A diagonal matrix from a vector $\mathbf{x} = [x_1, \ldots, x_N]^\mathrm{T}$ and a block diagonal matrix from matrices $\mathbf{X}_1, \ldots, \mathbf{X}_N$ are represented as $\mathrm{diag}(\mathbf{x})$ and $\mathrm{blkdiag}(\mathbf{X}_1, \ldots \mathbf{X}_N)$, respectively.
The notation $[\mathbf{A}]_{i,j}$ and $[\mathbf{A}]_{:,j}$ denote the $(i,j)$ element and the $j$-th column vector of the matrix $\mathbf{A}$, respectively.
The operators $\odot$, $\oslash$, and $\otimes$ represent the Hadamard product (element-wise multiplication), Hadamard division (element-wise division), and Kronecker product, respectively.
Given matrices $\mathbf{A} = [\mathbf{a}_1, \ldots, \mathbf{a}_N]$ and $\mathbf{B} = [\mathbf{b}_1, \ldots, \mathbf{b}_N]$, 
the notation $\mathbf{A} \circ \mathbf{B} \triangleq [\mathbf{a}_1 \otimes \mathbf{b}_1, \ldots, \mathbf{a}_N \otimes \mathbf{b}_N$] denotes the Khatri-Rao product between $\mathbf{A}$ and $\mathbf{B}$.
The notation $\| \mathbf{a} \|_p$ and $\|\mathbf{A}\|_\mathrm{F}$ indicate $\ell_p$-norm and Frobenius norm, respectively.
The notation $\mathrm{vec}(\mathbf{A})$ denotes the vectorization of the matrix $\mathbf{A}$ defined as $\mathrm{vec}(\mathbf{A}) = [\mathbf{a}_1^\mathrm{T}, \ldots, \mathbf{a}_N^\mathrm{T}]^\mathrm{T}$.
A circularly symmetric complex Gaussian distribution with mean $\bm{\mu}$ and covariance $\mathbf{C}$ is denoted as $\mathcal{CN}(\bm{\mu}, \mathbf{C})$.
The element-wise square root of the matrix $\mathbf{A}$ is denoted as $\sqrt{\mathbf{A}}$.

\section{System Model}
\label{sec:system_model}
    
% Fig: Pilot_allocation  
\begin{figure}[t!]
    \centering
    \includegraphics[width=\linewidth]{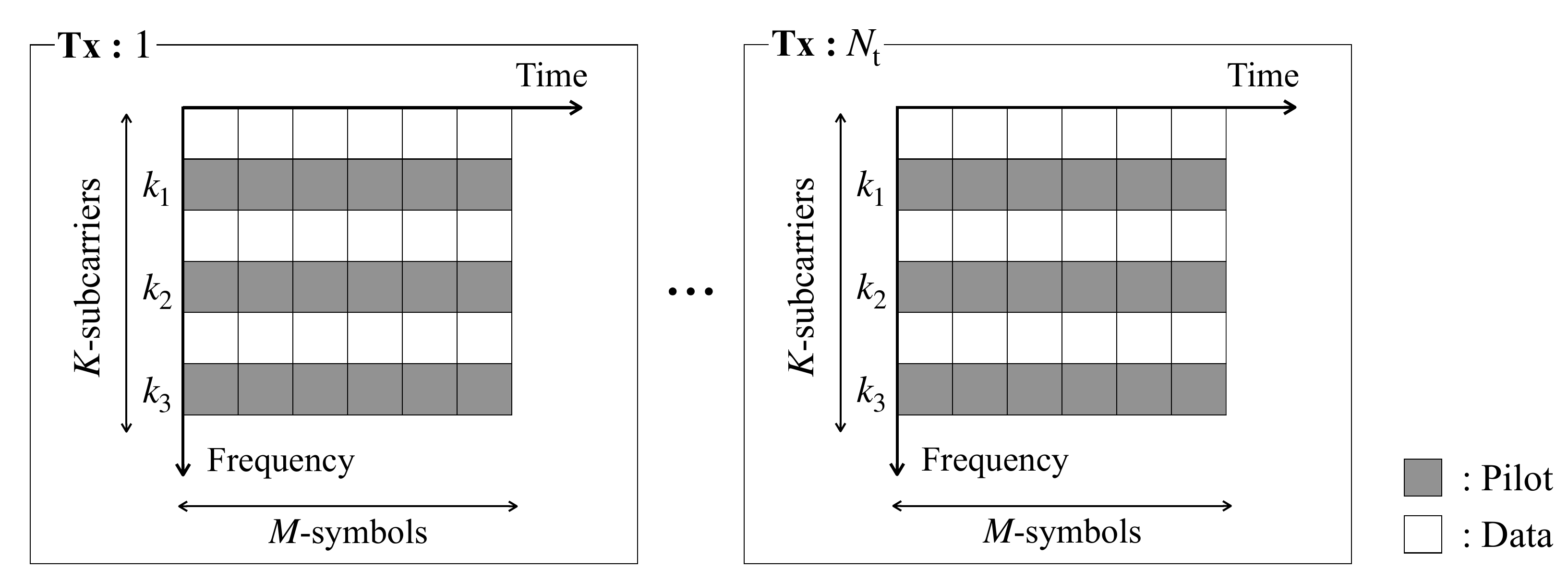}
    \caption{Pilot allocation in the time-frequency domain.}
    \label{fig:pilot_allocation}
\end{figure}

% system model 
We consider a downlink \ac{MIMO}-\ac{OFDM} system consisting of a \ac{BS} and multiple \acp{UE}.
The BS and UEs have $N_\mathrm{t}$ and $N_\mathrm{r}$-antenna \acp{ULA}, and the antenna spacing at the BS and the UE are $d_\mathrm{t}$ and $d_\mathrm{r}$, respectively.
The carrier frequency is denoted by $f_c$ and the carrier wavelength is expressed as $\lambda_\mathrm{c} = \frac{c}{f_\mathrm{c}}$ with the speed of light $c$. 
The system bandwidth and the number of subcarriers are defined as
$B$ and $K$, respectively.
The subcarrier index set is defined as $\mathcal{K} \triangleq \{1, 2, \ldots, K\}$.
Each \ac{OFDM} symbol is composed of $K$ subcarriers and the \ac{CP} with $N_\mathrm{cp}$ length.
The frequency corresponding to the $k$-th subcarrier $f_k$ is given by $f_k = f_\mathrm{c} +\Delta f_k$ with $\Delta f_k  = -B/2 + (k-1)B/K,\ k \in \mathcal{K}$.

For channel estimation between the BS and UEs, the BS transmits pilot signals to multiple UEs through downlink. 
Then, each UE estimates the channel using these pilot signals. 
Since all UEs perform the same procedure for channel estimation, this paper focuses only on a single UE without loss of generality.

\subsection{Channel Model}
We consider a frequency-selective MIMO channel model~\cite{2023Liu_CE_structured_MP}. 
The MIMO channel consists of $L$ resolvable delay paths with 
\acp{AoA} $\bm{\theta} \triangleq \{\theta_l\}_{l=1}^L$, 
\acp{AoD} $\bm{\phi} \triangleq \{\phi_l\}_{l=1}^L$, 
delay times $\bm{\tau} \triangleq \{\tau_l\}_{l=1}^L$, and 
path gains $\bm{\alpha} \triangleq [ \alpha_1, \alpha_2, \ldots, \alpha_L ]^\mathrm{T} \in \mathbb{C}^{L \times 1} $.
Then, the channel matrix between the BS and the UE at the $k$-th subcarrier $\mathbf{H}_k \in \mathbb{C}^{N_\mathrm{r} \times N_\mathrm{t}}$ is expressed as
\begin{align}
    \label{eq:H_k}
    \mathbf{H}_k &= \sum_{l=1}^L \alpha_l e^{-j 2 \pi \Delta f_k \tau_l} \ 
    \mathbf{a}_\mathrm{r}(\theta_l)
    \mathbf{a}_\mathrm{t}(\phi_l)^\mathrm{H},
\end{align}
where 
$\mathbf{a}_\mathrm{r} (\theta_{l}) \in \mathbb{C}^{N_\mathrm{r} \times 1}$, and 
$\mathbf{a}_\mathrm{t} (\phi_{l}) \in \mathbb{C}^{N_\mathrm{t} \times 1}$
are the array response vectors of the AoA and AoD.
For a given path, $l$, the array response vectors of AoA, AoD, and the delay response vector are defined as~\cite{2023Liu_CE_structured_MP}
\begin{align*}
    % a(AoA)
    \mathbf{a}_\mathrm{r} (\theta_{l}) &\triangleq \begin{bmatrix} 1, e^{j\frac{2 \pi}{\lambda_\mathrm{c}} d_\mathrm{r} \sin \theta_l}, \ldots, e^{j\frac{2 \pi}{\lambda_\mathrm{c}} (N_\mathrm{r} - 1) d_\mathrm{r} \sin \theta_l} \end{bmatrix}^\mathrm{T} \in \mathbb{C}^{N_\mathrm{r} \times 1}, \\
    % a(AoD)
    \mathbf{a}_\mathrm{t} (\phi_{l}) &\triangleq \begin{bmatrix} 1, e^{j\frac{2 \pi}{\lambda_\mathrm{c}} d_\mathrm{t} \sin \phi_l}, \ldots, e^{j\frac{2 \pi}{\lambda_\mathrm{c}} (N_\mathrm{t} - 1) d_\mathrm{t} \sin \phi_l} \end{bmatrix}^\mathrm{T}  \in \mathbb{C}^{N_\mathrm{t} \times 1}, \\
    % b(delay)
    \mathbf{b} (\tau_l) & \triangleq \begin{bmatrix} e^{-j 2 \pi \Delta f_1 \tau_l}, \ldots, e^{-j 2 \pi \Delta f_K \tau_l} \end{bmatrix}^\mathrm{T} \in \mathbb{C}^{K \times 1}.
\end{align*}

Stacking these vectors for $L$ paths, the array response matrices of AoA and AoD, and the delay response matrix are, respectively, defined as
\begin{subequations}
\begin{align}
    \label{eq:Ar}
    \mathbf{A}_\mathrm{r} (\bm{\theta}) &= \begin{bmatrix} \mathbf{a}_\mathrm{r} (\theta_1), \ldots, \mathbf{a}_\mathrm{r} (\theta_L) \end{bmatrix} \in \mathbb{C}^{N_\mathrm{r} \times L}, \\
    \label{eq:At}
    \mathbf{A}_\mathrm{t} (\bm{\phi}) &= \begin{bmatrix} \mathbf{a}_\mathrm{t} (\phi_1), \ldots, \mathbf{a}_\mathrm{t} (\phi_L) \end{bmatrix} \in \mathbb{C}^{N_\mathrm{t} \times L}, \\
    \label{eq:B}
    \mathbf{B} (\bm{\tau}) &= \begin{bmatrix} \mathbf{b} (\tau_1), \ldots, \mathbf{b} (\tau_L) \end{bmatrix} \in \mathbb{C}^{K \times L}.
\end{align}
\end{subequations}

For notational convenience, 
the $k$-th row vector of the delay response matrix $\mathbf{B} (\bm{\tau})$ in \eqref{eq:B}, denoted by $\bar{\mathbf{b}}_k(\mathbf{\tau}) \triangleq [\mathbf{B}(\bm{\tau})]_{k,:}^\mathrm{T} \in \mathbb{C}^{L \times 1}$, is expressed as 
\begin{align}
    \label{eq:b_tilde}
    \bar{\mathbf{b}}_k(\bm{\tau}) = \begin{bmatrix} e^{-j 2 \pi \Delta f_k \tau_1}, \ldots, e^{-j 2 \pi \Delta f_k \tau_L} \end{bmatrix}^\mathrm{T}.
\end{align}

From the matrices and vectors in \eqref{eq:Ar}--\eqref{eq:b_tilde}, the channel matrix in \eqref{eq:H_k} can be rewritten as
\begin{align}
    \label{eq:H_k_mat}
    \mathbf{H}_k &= \mathbf{A}_\mathrm{r} (\bm{\theta}) \mathrm{diag}\left( \bm{\alpha} \odot \bar{\mathbf{b}}_k(\bm{\tau}) \right) 
    \mathbf{A}_\mathrm{t}^\mathrm{H} (\bm{\phi}).
\end{align}

Using the vectorization property $ \mathrm{vec}(\mathbf{A} \mathrm{diag}(\mathbf{b}) \mathbf{C}) = (\mathbf{C}^\mathrm{T} \circ \mathbf{A}) \mathrm{vec}(\mathbf{b})$, the vectorized channel $\mathbf{h}_k \triangleq \mathrm{vec} (\mathbf{H}_k) \in \mathbb{C}^{N_\mathrm{r} N_\mathrm{t} \times 1}$ is expressed as
\begin{align}
    \label{eq:vec_h_k}
    \mathbf{h}_k = \left( \mathbf{A}_\mathrm{t}^\ast (\bm{\phi}) \circ \mathbf{A}_\mathrm{r} (\bm{\theta}) \right) (\bm{\alpha} \odot \bar{\mathbf{b}}_k(\bm{\tau})).
\end{align}

Stacking the vectorized channel $\mathbf{h}_k$ over the subcarrier direction $k \in \{1, 2, \ldots, K \}$, 
the channel matrix $\mathbf{H} \triangleq \begin{bmatrix} \mathbf{h}_1, \ldots, \mathbf{h}_K \end{bmatrix} \in \mathbb{C}^{N_\mathrm{r} N_\mathrm{t} \times K}$ is given by 
\begin{align}
    \label{eq:H}
    \mathbf{H} = \left( \mathbf{A}_\mathrm{t}^\ast (\bm{\phi}) \circ \mathbf{A}_\mathrm{r} (\bm{\theta}) \right) 
    \mathrm{diag} (\bm{\alpha})
    \mathbf{B}(\bm{\tau})^\mathrm{T}.
\end{align}

Utilizing the vectorization property again in \eqref{eq:H} yields the channel vector $\mathbf{h} \triangleq \mathrm{vec}(\mathbf{H}) \in \mathbb{C}^{N_\mathrm{r} N_\mathrm{t} K \times 1}$ as
\begin{align}
    \label{eq:h_vec}
    \mathbf{h} = \left( \mathbf{B}(\boldsymbol{\tau}) \circ
    \mathbf{A}_\mathrm{t}^\ast (\boldsymbol{\phi}) \circ
    \mathbf{A}_\mathrm{r}(\boldsymbol{\theta}) \right) \boldsymbol{\alpha}.
\end{align}

\subsection{Received Signal Model}
\label{subsec:rx_signal_mode}

For the channel estimation at the UE side through downlink, the BS transmits pilot sequences with symbol length $M$.
To reduce the pilot overhead for channel estimation, only $Q (\leq K)$ subcarriers from the total $K$ subcarriers are used for the pilots as shown in Fig.~\ref{fig:pilot_allocation}.
The subcarrier index set for pilots is denoted by $\mathcal{Q} \triangleq \{k_1, k_2 \ldots, k_{Q} \} \subset \mathcal{K}$.
The cardinality of the subcarrier set for pilots is equivalent to the number of pilot subcarriers as $|\mathcal{Q}|=Q$.
As is the case with standard reference signaling, such as 5G NR~\cite{3GPP_TS_38211}, Tx antennas do not overlay pilot symbols on data symbols in the same time-frequency resource elements as shown in Fig.~\ref{fig:pilot_allocation}.

In light of the above, let $x_{k,n_\mathrm{t},m}$ denote the pilot symbol at the $k$-th subcarrier, the $n_\mathrm{t}$-th BS antenna, and the $m$-th symbol.
The $n_\mathrm{t}$-th BS antenna transmits the pilot sequence 
corresponding to the $k$-th subcarrier
$\mathbf{x}_{k,n_\mathrm{t}} \triangleq \begin{bmatrix} x_{k, n_\mathrm{t}, 1}, \ldots, x_{k, n_\mathrm{t}, M} \end{bmatrix}^\mathrm{T} \in \mathbb{C}^{M \times 1}$ to the UE through downlink channels.
The received pilot matrix at the $k$-th subcarrier $\mathbf{Y}_k \in \mathbb{C}^{N_\mathrm{r} \times M}$ is expressed as
\begin{align}
    \label{eq:Y_k}
    \mathbf{Y}_k = \mathbf{H}_k \mathbf{X}_k + \mathbf{N}_k,
\end{align}
where $\mathbf{X}_k \triangleq \begin{bmatrix}
    \mathbf{x}_{k,1}, \ldots, \mathbf{x}_{k,N_\mathrm{t}}
\end{bmatrix}^\mathrm{T} \in \mathbb{C}^{N_\mathrm{t} \times M}$
is the pilot sequence matrix, and 
% $\mathbf{N}_k \in \mathbb{C}^{N \times K_\mathrm{p}}$ 
$\mathbf{N}_k \in \mathbb{C}^{N_\mathrm{r} \times M}$ 
is the \ac{AWGN} matrix, 
whose entries follow independent and identically distributed~(i.i.d.) $\mathcal{CN}(0, \sigma^2)$, where $\sigma^2$ is the noise variance.
The total transmit power for pilot sequences is denoted as
$P_\mathrm{t} \triangleq \sum_{k_q \in \mathcal{Q}} \| \mathbf{X}_{k_q} \|_\mathrm{F}^2$.
 
When the pilot sequences are orthogonal such that $\mathbf{x}_{k,n_\mathrm{t}}^\mathrm{H} \mathbf{x}_{k,n_\mathrm{t}^\prime} = 0 \ (n_\mathrm{t} \neq n_{\mathrm{t}^\prime})$, the received signal can be separated for each BS antenna. 
However, the orthogonal pilot requires $M \geq N_\mathrm{t}$, leading to large pilot overhead especially in large array systems.
Therefore, we utilize non-orthogonal pilot sequences such that $M < N_\mathrm{t}$ and $\mathbf{x}_{k,n_\mathrm{t}}^\mathrm{H} \mathbf{x}_{k,n_\mathrm{t}^\prime} \neq 0\ (n_\mathrm{t} \neq n_\mathrm{t}^\prime)$ to reduce pilot overhead as in~\cite{2019Stoica_frame_NOMA_Access, 2022Iimori_bilinear_Grant_free, 2018Stoica_frame_VTC, 2019Stoica_frame_mmWave_Acess}.

From \eqref{eq:Y_k}, the vectorized received pilot at the $k$-th subcarrier $\mathbf{y}_k \triangleq \mathrm{vec} (\mathbf{Y}_k) \in \mathbb{C}^{N_\mathrm{r} M \times 1}$ can be expressed as 
\begin{align}
    \mathbf{y}_k = (\mathbf{X}_k^\mathrm{T} \otimes \mathbf{I}_{N_\mathrm{r}}) \mathbf{h}_k + \mathbf{n}_k, 
\end{align}
where $\mathbf{h}_k$ is the vectorized channel in \eqref{eq:vec_h_k} and $\mathbf{n}_k \triangleq \mathrm{vec} (\mathbf{N}_k) \in \mathbb{C}^{N_r M \times 1}$ is the vectorized noise.

Let us define the channel vector on the subcarriers used for pilots $\mathcal{Q}$ as
$\mathbf{h}_\mathcal{Q} \triangleq [\mathbf{h}_{k_1}^\mathrm{T}, \ldots , \mathbf{h}_{k_Q}^\mathrm{T}]^\mathrm{T} \in \mathbb{C}^{N_\mathrm{r} N_\mathrm{t}Q \times 1}$.
Since $\mathbf{h}_\mathcal{Q}$ is a sub-vector of $\mathbf{h}$ in \eqref{eq:h_vec}, $\mathbf{h}_\mathcal{Q}$ can be expressed as 
\begin{align}
    \label{eq:h_Q}
    \mathbf{h}_\mathcal{Q} = \mathbf{S} \mathbf{h}, 
\end{align}
where
$\mathbf{S} \in \{0, 1\}^{N_\mathrm{r} N_\mathrm{t} Q \times N_\mathrm{r} N_\mathrm{t} K}$
is the binary mapping matrix, calculated as $\mathbf{S} = [\mathbf{I}_K]_{\mathcal{Q},:} \otimes \mathbf{I}_{N_\mathrm{r} N_\mathrm{t}}$.

Stacking the received pilot $\mathbf{y}_{k_q}$ over the pilot subcarrier direction $k_q \in \{k_1, k_2, \ldots, k_Q \}$, the stacked received pilot $\mathbf{y} \triangleq \begin{bmatrix} \mathbf{y}_{k_1}^\mathrm{T}, \ldots, \mathbf{y}_{k_Q}^\mathrm{T} \end{bmatrix}^\mathrm{T} \in \mathbb{C}^{N_\mathrm{r} M Q \times 1}$ can be expressed as 
\begin{align}
    \label{eq:y_vec_Q}
    \mathbf{y} &= \mathbf{D} \mathbf{h}_\mathcal{Q} + \mathbf{n}, \\
    \label{eq:y_vec}
    & = \mathbf{D} \mathbf{S} \mathbf{h} + \mathbf{n},
\end{align}
where 
$\mathbf{n} \triangleq [\mathbf{n}_{k_1}^\mathrm{T}, \ldots, \mathbf{n}_{k_Q}^\mathrm{T}]^\mathrm{T} \in \mathbb{C}^{N_\mathrm{r}MQ \times 1}$ is the noise vector.

The matrix $\mathbf{D}$ in \eqref{eq:y_vec} is expressed as $\mathbf{D} \triangleq \mathbf{X}_\mathrm{blk}^\mathrm{T} \otimes \mathbf{I}_{N_\mathrm{r}} \in \mathbb{C}^{N_\mathrm{r} M Q \times N_\mathrm{r} N_\mathrm{t} Q}$, where 
$\mathbf{X}_\mathrm{blk} \triangleq \mathrm{blkdiag}(\mathbf{X}_{k_1}, \ldots, \mathbf{X}_{k_Q}) \in \mathbb{C}^{N_\mathrm{t}Q \times MQ}$
is the pilot matrix consisting of $Q$ subcarriers.

Our objective is to estimate the channel vector $\mathbf{h}$ including all $K$ subcarriers, instead of $\mathbf{h}_\mathcal{Q}$, from the received signal $\mathbf{y}$ in \eqref{eq:y_vec} 
using the reduced number of pilot subcarriers~$Q(<K)$ and the reduced symbol length~$M(<N_\mathrm{t})$ to minimize pilot overhead.
A comprehensive description of the channel estimation is presented in the next section.

\section{Channel Estimation Exploiting Angle and Delay Sparsity}
\label{sec:channel_est}

% Problem in LS
In order to accurately estimate the channel using the measurement equation \eqref{eq:y_vec_Q} with classical channel estimation methods, such as \ac{LS} and \ac{LMMSE}, the number of pilot subcarriers $Q$ and pilot length $M$ must satisfy $Q=K$ and $M \ge N_\mathrm{t}$.
This requirement results in an increase in pilot overhead as $K$ and $N_\mathrm{t}$ grow.
% 
% CS-based
To reduce the pilot overhead for channel estimation, channel estimation methods exploiting channel sparsity in the delay-angle domain have been proposed in~\cite{2017Venugopal_CE_delay_hybrid, 2018Mo_CE_ADC, 2018Wang_CS_CE, 2019_CE_delay_turbo, 2023Liu_CE_structured_MP, 2024Uchimura_Tracking}, based on \ac{CS} frameworks.
These approaches enable efficient channel estimation even in the under-determined case of $Q < K$ and $M<N_\mathrm{t}$.

In CS-based estimation methods, virtual channel representation is commonly used, where the dictionary matrix is designed with quantized angle and delay grid points to obtain the sparse representation of the channel in the delay-angle domain.
The grids for \acp{AoA}, \acp{AoD} and delays are designed by quantizing the delay-angle domain into $G_\theta$, $G_\phi$, and $G_\tau$ points as
$\tilde{\bm{\theta}} = \{ \tilde{\theta}_{g_\theta} | g_\theta \in \{1,2, \ldots, G_\theta \}\}$, 
$\tilde{\bm{\phi}} = \{ \tilde{\phi}_{g_\phi} | g_\phi \in \{1,2, \ldots, G_\phi \}\}$, and 
$\tilde{\bm{\tau}} = \{ \tilde{\tau}_{g_\tau} | g_\tau \in \{1,2, \ldots, G_\tau \}\}$ 
with 
$\tilde{\theta}_{g_\theta} \in [-\pi/2, \pi/2]$,
$\tilde{\phi}_{g_\phi} \in [-\pi/2, \pi/2]$, and
$\tilde{\tau}_{g_\tau} \in [0,\ \tau_\mathrm{max}]$, respectively,
where $\tau_\text{max}$ represents the maximum delay time.

Using the angle and delay grids $\tilde{\bm{\theta}}$, $\tilde{\bm{\phi}}$, and $\tilde{\bm{\tau}}$, 
the dictionary matrix for AoA, AoD, and delay can be designed as 
\begin{subequations}
\begin{align}
    % Ar
    \label{eq:Ar_dict}
    \mathbf{A}_\mathrm{r}(\tilde{\bm{\theta}}) &\triangleq \begin{bmatrix}
        \mathbf{a}_\mathrm{r} (\tilde{\theta}_1), \ldots, \mathbf{a}_\mathrm{r} (\tilde{\theta}_{G_\theta})
    \end{bmatrix} \in \mathbb{C}^{N_\mathrm{r} \times G_\theta}, \\
    % At
    \label{eq:At_dict}
    \mathbf{A}_\mathrm{t}(\tilde{\bm{\phi}}) &\triangleq \begin{bmatrix}
        \mathbf{a}_\mathrm{t} (\tilde{\phi}_1), \ldots, \mathbf{a}_\mathrm{t} (\tilde{\phi}_{G_\phi})
    \end{bmatrix} \in \mathbb{C}^{N_\mathrm{t} \times G_\phi}, \\
    % B
    \label{eq:B_dict}
    \mathbf{B} (\tilde{\bm{\tau}}) &\triangleq \begin{bmatrix}
        \mathbf{b} (\tilde{\tau}_1), \ldots, \mathbf{b} (\tilde{\tau}_{G_\tau})
    \end{bmatrix} \in \mathbb{C}^{K \times G_\tau}.
\end{align}
\end{subequations}

Then, using the result from \eqref{eq:h_vec}, the channel vector $\mathbf{h}$ can be approximately expressed with the dictionaries $\mathbf{A}_\mathrm{r}(\tilde{\bm{\theta}})$, 
$\mathbf{A}_\mathrm{t}(\tilde{\bm{\phi}})$, and 
$\mathbf{B}(\tilde{\bm{\tau}})$ as
\begin{align}
    \label{eq:h_apx}
    \mathbf{h} \simeq
    \left(
    \mathbf{B}(\tilde{\boldsymbol{\tau}}) \otimes
    \mathbf{A}_\mathrm{t}^\ast (\tilde{\boldsymbol{\phi}}) \otimes
    \mathbf{A}_\mathrm{r}(\tilde{\boldsymbol{\theta}})
    \right) \tilde{\boldsymbol{\alpha}}, 
\end{align}
where $\tilde{\bm{\alpha}} \in \mathbb{C}^{G_\theta G_\phi G_\tau \times 1}$ is the virtual sparse path gain vector, and the number of nonzero elements of $\tilde{\bm{\alpha}}$ is denoted by $\hat{L}$.

From the virtual channel representation in \eqref{eq:h_apx}, 
the received pilot $\mathbf{y}$ in \eqref{eq:y_vec} can be expressed as
\begin{align}
    \label{eq:y_cs}
    \mathbf{y} \simeq \mathbf{\Psi}(\tilde{\boldsymbol{\tau}}, \tilde{\boldsymbol{\phi}}, \tilde{\boldsymbol{\theta}}) \tilde{\boldsymbol{\alpha}} + \mathbf{n}, 
\end{align}
where
$\mathbf{\Psi}(\tilde{\boldsymbol{\tau}}, \tilde{\boldsymbol{\phi}}, \tilde{\boldsymbol{\theta}}) \in \mathbb{C}^{N_\mathrm{r}MQ \times G_{\tau} G_{\phi} G_{\theta}} $
is the sensing matrix defined as
\begin{align}
    \label{eq:sensing_matrix}
    \mathbf{\Psi}(\tilde{\boldsymbol{\tau}}, \tilde{\boldsymbol{\phi}}, \tilde{\boldsymbol{\theta}}) \triangleq \mathbf{D} \mathbf{S} \left( \mathbf{B}(\tilde{\boldsymbol{\tau}}) \otimes \mathbf{A}_\mathrm{t}^\ast (\tilde{\boldsymbol{\phi}}) \otimes \mathbf{A}_\mathrm{r}(\tilde{\boldsymbol{\theta}}) \right).
\end{align}

For notation convenience, 
the total number of observations, including all UE antennas, symbols, and pilot subcarriers, is defined as $N \triangleq N_\mathrm{r} M Q$. 
The total number of grids, consisting of three one-dimensional grids with one for AoA, AoD, and delay, is defined as $G \triangleq G_\theta G_\phi G_\tau$.
Due to the large size of the sensing matrix $\mathbf{\Psi} \in \mathbb{C}^{N \times G}$, the computational burden in the design of the sensing matrix is one of the major challenges, as will be discussed in Section~\ref{subsec:efficient_comp}.

Based on the measurement equation in \eqref{eq:y_cs}, the optimization problem to estimate the sparse path gain vector $\tilde{\bm{\alpha}}$ is formulated as 
% Opt.
\begin{subequations}
\label{eq:CS}
\begin{align}
    \underset{\tilde{\bm{\alpha}}}{\mathrm{minimize}} \ &\left \| \tilde{\bm{\alpha}} \right \|_1, \\
    \mathrm{subject \ to}\ 
    &\left \| 
        \mathbf{y} - \mathbf{\Psi}(\tilde{\boldsymbol{\tau}}, \tilde{\boldsymbol{\phi}}, \tilde{\boldsymbol{\theta}}) \tilde{\boldsymbol{\alpha}} 
    \right \|_2 \leq \epsilon,
\end{align}
\end{subequations}
where $\epsilon$ is a parameter that mitigates overfitting to noisy observations.

The sparse reconstruction problem in \eqref{eq:CS} can be solved by CS-based techniques such as \ac{OMP}~\cite{1993Pati_OMP} and \ac{SBL}~\cite{2001Tipping_SBL}.
Let $\check{\bm{\alpha}} \in \mathbb{C}^{G \times 1}$ denote the $\hat{L}$-sparse vector obtained by solving problem \eqref{eq:CS} using a CS-based method, and $\hat{\bm{\alpha}} \in \mathbb{C}^{\hat{L} \times 1}$ denote a vector constructed by extracting the $\hat{L}$ nonzero elements from $\check{\bm{\alpha}}$.
The estimates of AoAs, AoDs, and delays, corresponding to the nonzero elements of $\check{\bm{\alpha}}$, are defined as $\hat{\boldsymbol{\theta}} \in \mathbb{R}^{\hat{L} \times 1}$, $\hat{\boldsymbol{\phi}} \in \mathbb{R}^{\hat{L} \times 1}$, and $\hat{\boldsymbol{\tau}} \in \mathbb{R}^{\hat{L} \times 1}$, respectively.
Then, the estimated channel vector can be reconstructed with $\hat{\boldsymbol{\theta}}$, $\hat{\boldsymbol{\phi}}$, $\hat{\boldsymbol{\tau}}$, and $\hat{\bm{\alpha}}$ as 
\begin{align}
    \label{eq:h_vec_est}
    \hat{\mathbf{h}} = \left( \mathbf{B}(\hat{\boldsymbol{\tau}}) \circ
    \mathbf{A}_\mathrm{t}^\ast (\hat{\boldsymbol{\phi}}) \circ
    \mathbf{A}_\mathrm{r}(\hat{\boldsymbol{\theta}}) \right) \hat{\boldsymbol{\alpha}}.
\end{align}

% CS performance
The sparse recovery performance in CS approaches depends on the structure of the sensing matrix~\cite{2012Yonina_CS_book}. 
One of the primary measures for the sensing matrix is \ac{RIP}~\cite{2012Yonina_CS_book}. 
For any $\hat{L}$-sparse vector $\tilde{\bm{\alpha}}$, we say that the sensing matrix $\mathbf{\Psi}$ satisfies the RIP of order $\hat{L}$ if there exists a constant $\delta \in (0,1)$ such that 
\begin{align}
    \label{eq:RIP}
    (1 - \delta) \| \tilde{\bm{\alpha}} \|_2^2 \leq \| \mathbf{\Psi} \tilde{\bm{\alpha}} \|_2^2 \leq (1 + \delta) \| \tilde{\bm{\alpha}} \|_2^2.
\end{align}

The smallest constant value of $\delta$ is called the \ac{RIC}, which is defined as $\delta_{\hat{L}}$.
As the RIC decreases, the orthogonality of the sensing matrix improves, which in turn enhances CS performance.
If the sensing matrix $\mathbf{\Psi}$ is orthogonal, i.e., $\bm{\psi}_i^\mathrm{H} \bm{\psi}_j = 0 \ (i \neq j)$, the RIC is $\delta_{\hat{L}} = 0$, where $\bm{\psi}_i$ represents the $i$-th column vector of $\mathbf{\Psi}$.
Assuming that the RIC satisfies $\delta_{2\hat{L}} < \sqrt{2} -1$, the estimation error between the optimal solution $\hat{\bm{\alpha}}$ in \eqref{eq:CS} and the true sparse vector $\bm{\alpha}_\mathrm{t} \in \mathbb{C}^{G \times 1}$ is bounded as 
\begin{align}
    \label{eq:CS_bound}
    \| \check{\bm{\alpha}} - \bm{\alpha}_\mathrm{t} \|_1 \leq \epsilon \cdot C,
\end{align}
where $C = \frac{4 \sqrt{1 + \delta_{2\hat{L}}}}{(1-(1-\sqrt{2}) \delta_{2 \hat{L} })}$ is a constant value depending on the RIC~\cite{2012Yonina_CS_book}.
From \eqref{eq:CS_bound}, in the noiseless case with $\epsilon=0$, the use of a sensing matrix $\mathbf{\Psi}$ satisfying $\delta_{2\hat{L}} < \sqrt{2} -1$ enables the exact recovery of a sparse vector without estimation errors.
However, computing the RIC for a given sensing matrix $\mathbf{\Psi}$ is computationally intractable within a reasonable time, because it requires a brute-force search over all combinations of $\hat{L}$ columns from $\mathbf{\Psi}$.

Another measure for the sensing matrix is mutual coherence, which is widely used due to its computational tractability.
The mutual coherence, given a sensing matrix $\mathbf{\Psi} \in \mathbb{C}^{N \times G}$, is defined as 
\begin{align}
    \label{eq:MC}
    \mu(\mathbf{\Psi}) \triangleq \underset{1 \leq i\neq j \leq G}{\mathrm{max}} \ \frac{ |\boldsymbol{\psi}_i^\mathrm{H} \boldsymbol{\psi}_j |}{\|\boldsymbol{\psi}_i\|_2 \|\boldsymbol{\psi}_j\|_2}.
\end{align}

The lower bound of the mutual coherence known as Welch bound~\cite{1974Welch_bound} is provided as
\begin{align}
    \label{eq:Welch_bound}
    \sqrt{\frac{G - N}{N(G-1)}} \leq \mu(\mathbf{\Psi})  \leq 1.
\end{align}

% RIC, MC
The mutual coherence is related to the RIC as $\delta_2 = \mu(\mathbf{\Psi})$, and the RIC of order $\hat{L}$ is bounded with $\mu(\mathbf{\Psi})$ as~\cite{2011Duarte_CS_theory}
\begin{align}
    \label{eq:MC_bound}
    \delta_{\hat{L}} \leq \mu(\mathbf{\Psi}) (\hat{L}-1).
\end{align}

Furthermore, if the mutual coherence satisfies the following inequality, 
\begin{align}
    \label{eq:MC_recovery_cond}
    \mu (\mathbf{\Psi}) < \frac{1}{2 \hat{L} - 1},
\end{align}
the $\hat{L}$-sparse vector $\check{\bm{\alpha}}$ that satisfies $\mathbf{y} = \mathbf{\Psi} \check{\bm{\alpha}}$ in the noiseless case is the unique solution.

From the mutual coherence properties in \eqref{eq:MC_bound} and \eqref{eq:MC_recovery_cond}, it can be seen that minimizing the mutual coherence, instead of directly treating the RIC, is an effective approach for sensing matrix design to improve sparse recovery performance in terms of computational feasibility.

\section{Joint Pilot allocation and Sequence Design}
\label{sec:prop}

As described in the previous section, the structure of the sensing matrix affects the sparse recovery performance.
In the considered MIMO-OFDM system, the sensing matrix $\mathbf{\Psi}$ is composed of the pilot sequences with symbol length $M$ over $Q$ subcarriers, $\{ \mathbf{X}_{k_q} \}_{k_q \in \mathcal{Q}}$, as defined in \eqref{eq:sensing_matrix}.
In this section, we propose a joint design for subcarrier allocation $\mathcal{Q}$ and pilot sequences $\{ \mathbf{X}_{k_q} \}_{k_q \in \mathcal{Q}}$ to enhance sparse recovery performance. 

The problem formulation for pilot optimization and its efficient computation are described in Sections~\ref{subsec:formulation} and \ref{subsec:efficient_comp}, respectively.
In Section~\ref{subsec:solve_opt}, a gradient descent-based method to solve the optimization problem is presented.
The computational complexity of the proposed optimization is evaluated in Section~\ref{subsec:complexity}.

\subsection{Formulation of Pilot Optimization Problem}
\label{subsec:formulation}

While the mutual coherence can guarantee the CS performance theoretically, the mutual coherence quantifies only the worst similarity between the column vectors of the sensing matrix, neglecting its average similarity.
Therefore, instead of mutual coherence, the total coherence is widely used for sensing matrix design in~\cite{2019Xiao_JMCTC, 2016He_allocation_MIMO_OFDM, 2011Zelnik_SensingDesign_Imag, 2022Ge_BeamDesign}.
The total coherence considers the average similarity, which leads to the improved performance compared to using only the mutual coherence.
The total coherence is related to the Gram matrix of the sensing matrix.
Using the equivalent sensing matrix $\bar{\mathbf{\Psi}} \in \mathbb{C}^{N \times G}$ with all column vectors normalized such that $\bar{\bm{\psi}}_i \triangleq \bm{\psi}_i / \| \bm{\psi}_i \|_2 $, the Gram matrix of the equivalent sensing matrix is given by $\mathbf{G}_{\mathbf{\Psi}} \triangleq \bar{\mathbf{\Psi}}^\mathrm{H} \bar{\mathbf{\Psi}} \in \mathbb{C}^{G \times G}$.
Then, the total coherence is defined, with the Gram matrix $\mathbf{G}_{\mathbf{\Psi}}$, as
\begin{align}
    \label{eq:TC}
    \gamma (\mathbf{\Psi}) \triangleq \left \| \mathbf{G}_{\mathbf{\Psi}} - \mathbf{I}_G \right  \|_\mathrm{F} = \left \{ \sum_{1 \leq i \neq j \leq G} \left (
        \frac{|\bm{\psi}_i^\mathrm{H} \bm{\psi}_j|}{\|\bm{\psi}_i\|_2 \|\bm{\psi}_j\|_2}
        \right )^2 \right \}^{1/2}.
\end{align}

As defined in \eqref{eq:TC}, the total coherence quantifies the sum of the squared off-diagonal elements in the Gram matrix, while the mutual coherence quantifies the largest off-diagonal element as shown in \eqref{eq:MC}.
The studies~\cite{2019Xiao_JMCTC, 2016He_allocation_MIMO_OFDM}  have demonstrated that the use of the total coherence for pilot subcarrier allocation enhances channel estimation performance in CS-based methods.
%%%%%%%%%%%%%%%%%%

Based on these studies, we introduce the generalized coherence $\nu_p (\mathbf{\Psi})$, which encompasses both the total coherence in \eqref{eq:TC} and the mutual coherence in \eqref{eq:MC} as special cases, as 
\begin{align}
    \label{eq:GC}
    \nu_p (\mathbf{\Psi}) \triangleq 
    \left \{ \sum_{1 \leq i \neq j \leq G} \left (
    \frac{|\bm{\psi}_i^\mathrm{H} \bm{\psi}_j|}{\|\bm{\psi}_i\|_2 \|\bm{\psi}_j\|_2}
    \right )^p \right \}^{1/p},
\end{align}
which is equivalent to the mutual coherence $\mu(\mathbf{\Psi})$ when $p \rightarrow \infty$, and is equivalent to the total coherence $\gamma(\mathbf{\Psi})$ when $p=2$.
As $p$ increases, large inner products become dominant among inner products.
This paper utilizes the generalized coherence in \eqref{eq:GC} for subcarrier allocation and pilot sequence design.

Although many studies have proposed pilot allocation or sequence design using coherence metrics, a joint design of subcarrier allocation and pilot sequence has not yet been proposed.
One of the difficulties in joint design is that the optimization problem for the pilot design is formulated as a \ac{MINLP} because pilot allocation requires discrete variables, while sequence design requires continuous variables.

With the generalized coherence in \eqref{eq:GC}, the optimization problem of joint design for subcarrier allocation $\mathcal{Q}$ and sequences $\{\mathbf{X}_{k_q} \}_{k_q \in \mathcal{Q}}$ is formulated as
% 
% MIP
\begin{subequations}
    \label{eq:opt_int}
    \begin{align}
        \label{eq:opt_int_obj}
        \underset{ \mathcal{Q},\ \{\mathbf{X}_{k_q}\}_{k_q \in \mathcal{Q}} }{\mathrm{minimize}} \ \ & \nu_p (\mathbf{\Psi}) \\
        \label{eq:opt_int_Q}
        \mathrm{subject\ to} \ \ & \mathcal{Q} = \{k_1, k_2, \ldots k_Q \} \subset \{1, 2, \ldots, K\}, \\
        \label{eq:opt_int_P}
        & \sum_{k_q \in \mathcal{Q}} \| \mathbf{X}_{k_q} \|_\mathrm{F}^2 = P_\mathrm{t},
    \end{align}
\end{subequations}
where 
\eqref{eq:opt_int_Q} is the integer constraint for subcarrier indices in $\mathcal{Q}$, and 
\eqref{eq:opt_int_P} is the constraint for the total pilot transmit power $P_\mathrm{t}$.

As formulated in \eqref{eq:opt_int_obj}–\eqref{eq:opt_int_P}, the problem involves integer variables in $\mathcal{Q}$ and is therefore a \ac{MINLP}. Exact solution is challenging, since the number of possible subcarrier allocations grows combinatorially as $\frac{K!}{Q! (K-Q)!}$.  
For example, when $K=64,\ Q=10$, the number of candidates for the allocation of subcarriers is $1.51 \times 10^{11}$.
Therefore, it is impractical to solve the problem via brute-force search while simultaneously considering sequence design.

To efficiently deal with the optimization problem in \eqref{eq:opt_int}, the original problem that includes integer variables is transformed into the problem without integer variables.
In the proposed method, we design the pilot sequences over all $K$ subcarriers, $\{ \mathbf{X}_k \}_{k=1}^K$, instead of the limited subcarrier indices $\mathcal{Q} = \{k_1, \ldots, k_Q \}$.
Then, by introducing a block sparse penalty for the pilot sequences $\{ \mathbf{X}_k \}_{k=1}^K$, unnecessary pilot sequences whose contribution to the coherence metric is small are forced to zero among a total of $K$ subcarriers.
To elaborate, defining the pilot sequence matrix over all $K$ subcarriers as $\mathbf{X} \triangleq [\mathbf{X}_1, \ldots, \mathbf{X}_K] \in \mathbb{C}^{N_\mathrm{t} \times MK}$, the block sparse penalty for $\mathbf{X}$ is introduced as
% 
% Sparse penalty
\begin{align}
    \label{eq:penalty_sp}
    g(\mathbf{X}) \triangleq 
    \left ( 
        \sum_{k=1}^K  \| \mathbf{X}_k \|_{\mathrm{F}}^q
    \right )^{1/q} = \|\mathbf{v}(\mathbf{X}) \|_q, \  (0 < q \leq 1),
\end{align}
where
$\mathbf{v}(\mathbf{X}) \triangleq \begin{bmatrix} \| \mathbf{X}_1 \|_\mathrm{F}, \ldots, \| \mathbf{X}_K \|_\mathrm{F} \end{bmatrix}^\mathrm{T} \in \mathbb{R}^{K \times 1}$ is a vector whose $k$-th element represents the magnitude of the $k$-th subcarrier.
Introducing the block sparse penalty \eqref{eq:penalty_sp} in the optimization will promote sparse subcarrier allocation without requiring integer variables.

Since the generalized coherence in \eqref{eq:GC} consists of fractional components of the sensing matrix, making analytical tractability difficult, we transform the generalized coherence by introducing a penalty leading the $\ell_2$-norms of the column vectors approach the same value such that $\| \bm{\psi}_1 \|_2 \simeq \cdots \simeq \| \bm{\psi}_G \|_2$.
Let the transformed generalized coherence be denoted as $f^\mathbf{\Psi}(\mathbf{X})$ for an explicit expression as a function of the pilot matrix $\mathbf{X}$, and then $f^\mathbf{\Psi}(\mathbf{X})$ is formulated as 
\begin{align}
    \label{eq:fx}
    f^\mathbf{\Psi} (\mathbf{X})
    & \triangleq \left ( \sum_{1 \leq i,j \leq G} 
    |\bm{\psi}_i^\mathrm{H} \bm{\psi}_j |^p \right )^{1/p} \\
    % 
    % \label{eq:fx_decompose}
    &= \left ( \sum_{1 \leq i \neq j \leq G} 
    |\bm{\psi}_i^\mathrm{H} \bm{\psi}_j |^p 
    + \sum_{1 \leq i \leq G} \| \bm{\psi}_i \|_2^{2p} 
    \right )^{1/p},  \nonumber
\end{align}
where the first term quantifies the similarity between the column vectors of the sensing matrix, and the second term is a penalty for the $\ell_2$-norms of the column vectors.
Given a large value of $p$, the large inner products (i.e., $\ell_2$-norms for $i = j$, corresponding to the second term) are dominant in the objective function. 
Thus, the minimization of the transformed generalized coherence in \eqref{eq:fx} leads to $\ell_2$-norms approaching the same value such that $\| \bm{\psi}_1 \|_2 \simeq \cdots \simeq \| \bm{\psi}_G \|_2$.
Note that, owing to the power constraint in \eqref{eq:opt_int_P}, the solution obtained by minimizing $f^\mathbf{\Psi}(\mathbf{{X}})$ does not become the trivial solution $\bm{\psi}_i = \mathbf{0},\ \forall i \in \{1, \ldots, G \}$.

From the transformed generalized coherence $f^\mathbf{\Psi}(\mathbf{X})$ in \eqref{eq:GC} and the block sparse penalty for subcarrier allocation $g(\mathbf{X})$ in \eqref{eq:penalty_sp}, the optimization problem for joint pilot allocation and sequence design is formulated as
% 
% Opt. prop.
\begin{subequations}
    \label{eq:opt}
    \begin{align}
        \label{eq:opt_obj}
        \underset{ \mathbf{X} }{\mathrm{minimize}} \ \ & f^\mathbf{\Psi}(\mathbf{X}) + \lambda g(\mathbf{X}) \\
        \label{eq:opt_P}
        \mathrm{subject \ to} \ \ & \| \mathbf{X} \|_\mathrm{F}^2 = P_\mathrm{t},
    \end{align}
\end{subequations}
where $\lambda$ is a hyper-parameter to control the sparsity of $ \{ \mathbf{X}_k \}_{k=1}^K$.

\subsection{Efficient Computation of Generalized Coherence}
\label{subsec:efficient_comp}

As a key contribution of this paper, this section explores the efficient computation of the transformed generalized coherence in \eqref{eq:fx} by leveraging the structure of the sensing matrix.

As defined in \eqref{eq:sensing_matrix}, the size of the sensing matrix $\mathbf{\Psi}$ is significantly large since the column dimension of $\mathbf{\Psi}$ is $N = N_\mathrm{r} M K$, including the number of UE antennas $N_\mathrm{r}$, the length of sequences $M$, and the number of subcarriers $K$, and the row dimension of $\mathbf{\Psi}$ is $G = G_\theta G_\phi G_\tau$. 
Let $g_\tau,\ g_\phi$ and $g_\theta$ denote the indices for the delay, AoD, and AoA grids of the dictionary matrices $\mathbf{B} (\tilde{\bm{\tau}}), \ \mathbf{A}_\mathrm{t} (\tilde{\bm{\phi}})$, and $\mathbf{A}_\mathrm{r} (\tilde{\bm{\theta}})$, respectively.
Then, the column index of $\mathbf{\Psi}$ composed of these dictionaries, $g \in \{1,2,\ldots, G_\theta G_\phi G_\tau \}$, can be expressed as 
\begin{align}
    \label{eq:g}
    g = G_\theta G_\phi (g_\tau -1) + G_\theta (g_\phi - 1) + g_\theta,
\end{align}
with
$g_\tau \in \{1,2,\ldots, G_\tau\}$, $g_\phi \in \{1,2,\ldots, G_\phi \}$, and $g_\theta \in \{1,2,\ldots, G_\theta \}$.
From the sensing matrix $\mathbf{\Psi}$ in \eqref{eq:sensing_matrix}, the $g$-th column vector $\bm{\psi}_g \in \mathbb{C}^{N_\mathrm{r} MK \times 1}$ is given by
\begin{align}
    \label{eq:psi_g}
    \bm{\psi}_g = \mathbf{D} \mathbf{S} \left ( \mathbf{b} (\tilde{\tau}_{g_\tau}) 
    \otimes \mathbf{a}^\ast_{\mathrm{t}} (\tilde{\phi}_{g_{\phi}}) 
    \otimes \mathbf{a}_{\mathrm{r}} (\tilde{\theta}_{g_{\theta}}) 
    \right ),
\end{align}
where the binary mapping matrix $\mathbf{S}$ defined in \eqref{eq:h_Q} is given by $\mathbf{S} = \mathbf{I}_{N_\mathrm{r} N_\mathrm{t} K}$ since the proposed method designs the pilot sequence over all $K$ subcarriers rather than a limited subcarrier set $\mathcal{Q}$.

For efficient computation of the transformed generalized coherence in \eqref{eq:fx}, the sensing matrix in \eqref{eq:sensing_matrix} is rewritten, using the property of Kronecker products $(\mathbf{A} \otimes \mathbf{B}) (\mathbf{C} \otimes \mathbf{D}) = (\mathbf{AC}) \otimes (\mathbf{BD})$, as 
\begin{align}
    \mathbf{\Psi} &= (\mathbf{X}_\mathrm{blk}^\mathrm{T} \otimes \mathbf{I}_{N_\mathrm{r}})
    (\tilde{\mathbf{B}} \otimes \tilde{\mathbf{A}}_\mathrm{t}^\ast \otimes \tilde{\mathbf{A}}_\mathrm{r}) \\
    & = \underbrace{  
        \left \{ \mathbf{X}_\mathrm{blk}^\mathrm{T} (\tilde{\mathbf{B}} \otimes \tilde{\mathbf{A}}_\mathrm{t}^\ast) \right \} 
    }_{ \triangleq \mathbf{\Omega} \in \mathbb{C}^{N_\mathrm{t} K \times G_\tau G_\phi} }
     \otimes \tilde{\mathbf{A}}_\mathrm{r} 
    = \mathbf{\Omega} \otimes \tilde{\mathbf{A}}_\mathrm{r}, 
\end{align}
where $\mathbf{\Omega}$ is the part of the sensing matrix excluding the dictionary for AoA, $\mathbf{A}_\mathrm{r}(\tilde{\bm{\theta}})$.
For the $g_\tau$-th delay grid and the $g_\phi$-th AoD grid, the column vector of $\mathbf{\Omega}$ is given by 
\begin{align}
    \label{eq:omega_g}
    \bm{\omega}_{g_\tau,g_\phi} 
    = \mathbf{X}_\mathrm{blk}^\mathrm{T} \left (
    \mathbf{b}( \tilde{\tau}_{g_\tau}) \otimes \mathbf{a}_{\mathrm{t}}^\ast (\tilde{\phi}_{g_\phi})  \right ).
\end{align}

For notational convenience, let us define the inner product between the $g$-th and $g^\prime$-th column vectors for the transformed generalized coherence in \eqref{eq:fx} as $c^\mathbf{\Psi}_{g, g^\prime} (\mathbf{X}) \triangleq \bm{\psi}_g^\mathrm{H} \bm{\psi}_{g^\prime} \in \mathbb{C}$.
From \eqref{eq:psi_g}, \eqref{eq:omega_g}, the inner product $c^\mathbf{\Psi}_{g, g^\prime} (\mathbf{X})$ can be calculated as

\begin{align}
    \label{eq:c_psi_omega}
    c^\mathbf{\Psi}_{g, g^\prime} (\mathbf{X}) 
    &= \left ( \bm{\omega}^\mathrm{H}_{g_\tau,g_\phi} \otimes \mathbf{a}^\mathrm{H}_{\mathrm{r}} (\tilde{\theta}_{g_\theta}) \right ) 
    \left ( \bm{\omega}_{g^\prime_\tau,g^\prime_\phi} \otimes \mathbf{a}_{\mathrm{r}} (\tilde{\theta}_{g^\prime_\theta}) \right ) \nonumber \\
    &\overset{(a)}{=} \underbrace{ \left ( \bm{\omega}^\mathrm{H}_{g_\tau,g_\phi} \bm{\omega}_{g^\prime_\tau,g^\prime_\phi} \right )}_{\triangleq c^\mathbf{\Omega}_{g_\tau, g_\tau^\prime, g_\phi, g_\phi^\prime} (\mathbf{X}) \in \mathbb{C}}
    \left ( \mathbf{a}^\mathrm{H}_{\mathrm{r}} (\tilde{\theta}_{g_\theta}) \mathbf{a}_{\mathrm{r}} (\tilde{\theta}_{g^\prime_\theta}) \right ),
\end{align}
where the indices $g_\tau, g_\tau^\prime, g_\phi, g_\phi^\prime, g_\theta,g_\theta^\prime$ satisfy the relation in \eqref{eq:g}, and 
the equality of $(a)$ holds using the property of Kronecker products, $(\mathbf{A} \otimes \mathbf{B}) (\mathbf{C} \otimes \mathbf{D}) = (\mathbf{AC}) \otimes (\mathbf{BD})$.
In \eqref{eq:c_psi_omega}, the inner product related to the delay and AoD grids, $c^\mathbf{\Omega}_{g_\tau, g_\tau^\prime, g_\phi, g_\phi^\prime} (\mathbf{X})$, depends on the pilot sequence $\mathbf{X}$, while the inner product related to the AoD grids, $\left ( \mathbf{a}^\mathrm{H}_{\mathrm{r}} (\tilde{\theta}_{g_\theta}) \mathbf{a}_{\mathrm{r}} (\tilde{\theta}_{g^\prime_\theta}) \right )$, does not depend on $\mathbf{X}$.

Using this property from \eqref{eq:c_psi_omega}, the transformed generalized coherence $f^\mathbf{\Psi}(\mathbf{X})$ in \eqref{eq:fx} can be reformulated as
\begin{align}
    \label{eq:fx_coh}
    &f^\mathbf{\Psi}(\mathbf{X})
    = \left ( \sum_{1 \leq g,g^\prime \leq G} 
    | c^{\mathbf{\Psi}}_{g,g^\prime} (\mathbf{X}) |^p 
    \right )^{1/p} \nonumber \\
    & \! = \! \left ( \!
        \sum_{\substack{1 \leq g_\tau,g_\tau^\prime \leq G_\tau \\ 1 \leq g_\phi,g_\phi^\prime \leq G_\phi} } \! \! \! \! \!
        \left | c^\mathbf{\Omega}_{g_\tau, g_\tau^\prime, g_\phi, g_\phi^\prime} \! (\mathbf{X}) \right |^p \! \! \! \! \!
        \sum_{1 \leq g_\theta, g_\theta^\prime \leq G_\theta} 
        \! \! \! \! \left |
        \mathbf{a}^\mathrm{H}_{\mathrm{r}} (\tilde{\theta}_{g_\theta}) \mathbf{a}_{\mathrm{r}} (\tilde{\theta}_{g^\prime_\theta}) \right |^p \! \!
    \right )^{\! \! \! 1/p} \nonumber \\
    &= t_p \left (\mathbf{A}_\mathrm{r} (\tilde{\bm{\theta}}) \right ) \cdot 
    \underbrace{
    \left (
    \sum_{\substack{1 \leq g_\tau,g_\tau^\prime \leq G_\tau \\ 1 \leq g_\phi,g_\phi^\prime \leq G_\phi} } 
        \left | c^\mathbf{\Omega}_{g_\tau, g_\tau^\prime, g_\phi, g_\phi^\prime} (\mathbf{X}) \right |^p
    \right )^{1/p} }_{\triangleq f^\mathbf{\Omega}(\mathbf{X})},
\end{align}
where 
$t_p \left (\mathbf{A}_\mathrm{r} (\tilde{\bm{\theta}}) \right ) = 
\left ( \sum_{1 \leq g_\theta, g_\theta^\prime \leq G_\theta} \left | \mathbf{a}^\mathrm{H}_{\mathrm{r}} (\tilde{\theta}_{g_\theta}) \mathbf{a}_{\mathrm{r}} (\tilde{\theta}_{g^\prime_\theta}) \right |^p \right )^{1/p}$ 
is the coherence of the dictionary $\mathbf{A}_\mathrm{r} (\tilde{\bm{\theta}})$, which does not depend on the pilot matrix $\mathbf{X}$.
Therefore, the minimization of $f^\mathbf{\Psi} (\mathbf{X})$ is equivalent to the minimization of $f^\mathbf{\Omega}(\mathbf{X})$.

Defining the delay response vector for the $g_\tau$-th delay grid as 
$\mathbf{b} (\tilde{\tau}_{g_\tau}) \triangleq
\begin{bmatrix}
    b_{1} (\tilde{\tau}_{g_\tau}) \ldots, b_{K} (\tilde{\tau}_{g_\tau})
\end{bmatrix}^\mathrm{T}$,
the vector $\bm{\omega}_{g_\tau, g_\theta}$ can be rewritten from \eqref{eq:omega_g} as
\begin{align}
    \label{eq:omega_vec}
    \bm{\omega}_{g_\tau, g_\phi}
    =\begin{bmatrix}
        b_1 (\tilde{\tau}_{g_\tau}) \mathbf{a}_{\mathrm{t}}^\mathrm{H} (\tilde{\phi}_{g_\phi}) \mathbf{X}_1, \ldots, 
        b_K (\tilde{\tau}_{g_\tau}) \mathbf{a}_{\mathrm{t}}^\mathrm{H} (\tilde{\phi}_{g_\phi}) \mathbf{X}_K
    \end{bmatrix}^\mathrm{T}.
\end{align}

Thus, from \eqref{eq:omega_vec}, the inner product $c^\mathbf{\Omega}_{g_\tau, g_\tau^\prime, g_\phi, g_\phi^\prime} (\mathbf{X})$ in \eqref{eq:fx_coh} can be calculated as
\begin{align}
    \label{eq:c_omega_eff}
    &c^\mathbf{\Omega}_{g_\tau, g_\tau^\prime, g_\phi, g_\phi^\prime} (\mathbf{X}) \nonumber \\
    &=
    \mathbf{a}_{\mathrm{t}}^\mathrm{T} (\tilde{\phi}_{g_\phi})
    \left (
    \sum_{k=1}^K b_{k}^\ast (\tilde{\tau}_{g_\tau}) \mathbf{X}_k^\ast \mathbf{X}_k^\mathrm{T} 
    b_{k} (\tilde{\tau}_{g_\tau^\prime}) \right ) \mathbf{a}_{\mathrm{t}}^\ast (\tilde{\phi}_{g_\phi^\prime}).
\end{align}

Consequently, from \eqref{eq:fx_coh}, \eqref{eq:c_omega_eff}, the objective function can be expressed as \eqref{eq:fx_omega} at the top of the next page.

As illustrated above, through some mathematical operations in \eqref{eq:fx_coh}, \eqref{eq:c_omega_eff}, the objective function $f^\mathbf{\Omega}(\mathbf{X})$ in \eqref{eq:fx_omega} can be calculated such that it does not depend on the dictionary matrix for the AoA, while the original objective function $f^\mathbf{\Psi} (\mathbf{X})$ in \eqref{eq:fx} involves a large-sized dictionary matrix over the three-dimensional grid composed of the delay, AoA, and AoD. 
As a result, the computational complexity of $f^\mathbf{\Omega} (\mathbf{X})$ can be significantly reduced compared to the original objective function $f^\mathbf{\Psi} (\mathbf{X})$.
The detailed complexity analysis is provided in Section~\ref{subsec:complexity}

\begin{figure*}
\begin{align}
    \label{eq:fx_omega}
    f^\mathbf{\Omega}(\mathbf{X}) = 
    \left (
    \sum_{g_\tau,g_\tau^\prime, g_\phi,g_\phi^\prime}  
    \left |
    \mathbf{a}_{\mathrm{t}}^\mathrm{T} (\tilde{\phi}_{g_\phi})
    \left ( \sum_{k=1}^K b_{k}^\ast (\tilde{\tau}_{g_\tau}) \mathbf{X}_k^\ast \mathbf{X}_k^\mathrm{T} 
    b_{k} (\tilde{\tau}_{g_\tau^\prime}) \right ) \mathbf{a}_{\mathrm{t}}^\ast (\tilde{\phi}_{g_\phi^\prime}) \right |^p
    \right )^{1/p}.
\end{align}
\hrulefill
\end{figure*}

\subsection{Gradient Descent-Based Methods for Pilot Optimization}
\label{subsec:solve_opt}

Using the transformed generalized coherence $f^\mathbf{\Omega}(\mathbf{X})$ in \eqref{eq:fx_omega}, the joint optimization problem for pilot allocation and pilot sequence can be reformulated as
% 
% Opt. prop.
\begin{subequations}
    \label{eq:opt_omega}
    \begin{align}
        \label{eq:opt_obj_omega}
        \underset{ \mathbf{X} }{\mathrm{minimize}} \ \ & f^\mathbf{\Omega}(\mathbf{X}) + \lambda g(\mathbf{X}) \\
        \label{eq:opt_P_omega}
        \mathrm{subject \ to} \ \ & \| \mathbf{X} \|_\mathrm{F}^2 = P_\mathrm{t}.
    \end{align}
\end{subequations}

% Power constraint
To simplify the power constraint in \eqref{eq:opt_P_omega}, we introduce a new variable $\bar{\mathbf{X}} \in \mathbb{C}^{N_\mathrm{t} \times MK}$ that satisfies 
\begin{align}
    \label{eq:X_bar}
    \mathbf{X} = \frac{\sqrt{P_\mathrm{t}}}{\| \bar{\mathbf{X}} \|_\mathrm{F}} \bar{\mathbf{X}},
\end{align}
with 
$\bar{\mathbf{X}} = \begin{bmatrix} \bar{\mathbf{X}}_1, \ldots,\ \bar{\mathbf{X}}_K \end{bmatrix}$.

The use of this expression in \eqref{eq:X_bar} ensures that the pilot sequence always satisfies the power constraint in \eqref{eq:opt_P_omega}. 
Consequently, the optimization problem with the power constraint in \eqref{eq:opt_omega} can be transformed into an unconstrained optimization problem. 
By substituting \eqref{eq:X_bar} into \eqref{eq:opt_obj_omega}, the objective function $J(\mathbf{X})$ can be rewritten as 
\begin{align}
    \label{eq:obj_unconst}
    J(\mathbf{X}) 
    \triangleq f^\mathbf{\Omega}(\mathbf{X}) + \lambda g(\mathbf{X})
    = P_\mathrm{t} \cdot 
    \underbrace{
    \left ( \frac{f^\mathbf{\Omega} (\bar{\mathbf{X}})}{ \| \bar{\mathbf{X}}\|_\mathrm{F}^2}
    + \bar{\lambda} \frac{g(\bar{\mathbf{X}})}{\|\bar{\mathbf{X}}\|_\mathrm{F}} \right )
    }_{\triangleq L(\bar{\mathbf{X}})},
\end{align}
where $\bar{\lambda} = \lambda / \sqrt{P_\mathrm{t}}$ is a hyper-parameter that controls the sparsity of the pilot sequence $\bar{\mathbf{X}}$.
Since it is evident that the coefficient $P_\mathrm{t}$ in \eqref{eq:obj_unconst} does not affect the shape of the objective function $J(\mathbf{X})$, the minimization of $J(\mathbf{X})$ is equivalent to the minimization of $L(\bar{\mathbf{X}})$.
Hence, the unconstrained optimization problem for pilot design can be formulated as 
\begin{align}
    \label{eq:opt_unconst}
    \underset{\bar{\mathbf{X}}}{\mathrm{minimize}} \ \  
    \frac{f^\mathbf{\Omega}(\bar{\mathbf{X}})}{ \| \bar{\mathbf{X}}\|_\mathrm{F}^2}
    + \bar{\lambda}
    \frac{g(\bar{\mathbf{X}})}{\|\bar{\mathbf{X}}\|_\mathrm{F}},
\end{align}
where $f^\mathbf{\Omega}(\bar{\mathbf{X}})$ and $g(\bar{\mathbf{X}})$ are defined in \eqref{eq:fx_omega} and \eqref{eq:penalty_sp}.

Since the optimization problem in \eqref{eq:opt_unconst} is an unconstrained problem that includes only continuous variables without integer variables, it can be solved using gradient descent-based methods such as \ac{Adam}~\cite{2014kingma_Adam}. 

The gradient of the objective function, using the Wirtinger derivative~\cite{2007Hjorungnes_ComplexGrad}, is given by 
\begin{align}
    \label{eq:grad_L}
    \frac{\partial L(\bar{\mathbf{X}})}{\partial \bar{\mathbf{X}}_k^\ast} = 
    \frac{\partial}{\partial \bar{\mathbf{X}}_k^\ast} \left (
    \frac{f^\mathbf{\Omega}(\bar{\mathbf{X}})}{ \| \bar{\mathbf{X}}\|_\mathrm{F}^2} \right )
    + \bar{\lambda}
    \frac{\partial}{\partial \bar{\mathbf{X}}_k^\ast} \left (
    \frac{g(\bar{\mathbf{X}})}{\|\bar{\mathbf{X}}\|_\mathrm{F}} \right ).
\end{align}
% 
%%%%%%%%%%%%%%%%%%%%%%%%%%%%%%%%%%%%%%%%%
% Gradients
\begin{figure*}
\begin{align}
    \label{eq:grad_obj}
    \frac{\partial}{\partial \bar{\mathbf{X}}_k^\ast}
    \left (\frac{f^\mathbf{\Omega}(\bar{\mathbf{X}})}{ \| \bar{\mathbf{X}}\|_\mathrm{F}^2} \right )
    &= \frac{f^\mathbf{\Omega}(\bar{\mathbf{X}})}{\| \bar{\mathbf{X}} \|_\mathrm{F}^2}
    \left[
        \frac{-1}{\| \bar{\mathbf{X}} \|_\mathrm{F}^2} \mathbf{I}_{N_\mathrm{t}} + \frac{1}{ p v_p(\bar{\mathbf{X}})}
        \sum_{g_\tau, g_\phi, g_\tau^\prime, g_\phi^\prime} 
        \frac{p}{2} \left| c^\mathbf{\Omega}_{g_\tau, g_\tau^\prime, g_\phi, g_\phi^\prime} (\bar{\mathbf{X}}) \right|^{p-2}
        \left ( 
        \mathbf{V}_{g_\tau, g_\tau^\prime, g_\phi, g_\phi^\prime} + \mathbf{V}_{g_\tau, g_\tau^\prime, g_\phi, g_\phi^\prime}^\mathrm{H} 
        \right )
    \right] \bar{\mathbf{X}}_k, \\
    \label{eq:grad_sp}
    \frac{\partial}{\partial \bar{\mathbf{X}}_k^\ast} \left (
    \frac{g(\bar{\mathbf{X}})}{\|\bar{\mathbf{X}}\|_\mathrm{F}} \right )
    &= \frac{g(\bar{\mathbf{X}})}{2 \|\bar{\mathbf{X}}\|_\mathrm{F}} 
    \left[
    \frac{-1}{\| \bar{\mathbf{X}} \|_\mathrm{F}^2} + \frac{1}{\|\bar{\mathbf{X}}_k \|_\mathrm{F}^2} \left( \frac{\|\bar{\mathbf{X}}_k\|_\mathrm{F}}{g(\bar{\mathbf{X}})}\right)^{q}
    \right] \bar{\mathbf{X}}_k.
\end{align}
\hrulefill
\end{figure*}
%%%%%%%%%%%%%%%%%%%%%%%%%%%%%%%%%%%%%%%%%

In \eqref{eq:grad_L}, the gradients 
$\frac{\partial}{\partial \bar{\mathbf{X}}_k^\ast} \left (\frac{f^\mathbf{\Omega}(\bar{\mathbf{X}})}{ \| \bar{\mathbf{X}}\|_\mathrm{F}^2} \right )$ and 
$\frac{\partial}{\partial \bar{\mathbf{X}}_k^\ast} \left ( \frac{g(\bar{\mathbf{X}})}{\|\bar{\mathbf{X}}\|_\mathrm{F}} \right )$ 
can be calculated as \eqref{eq:grad_obj}, \eqref{eq:grad_sp} at the top of the next page, where $v_p(\bar{\mathbf{X}}) \in \mathbb{R}$ and $\mathbf{V}_{g_\tau, g_\tau^\prime, g_\phi, g_\phi^\prime} \in \mathbb{C}^{N_\mathrm{t} \times N_\mathrm{t}}$ are defined as
\begin{align}
    \label{eq:vp}
    v_p(\bar{\mathbf{X}}) 
    & \triangleq \sum_{g_\tau, g_\phi, g_\tau^\prime, g_\phi^\prime} \left | c^\mathbf{\Omega}_{g_\tau, g_\tau^\prime, g_\phi, g_\phi^\prime} (\bar{\mathbf{X}})\right |^p, \\
    \label{eq:V}
    \mathbf{V}_{g_\tau, g_\tau^\prime, g_\phi, g_\phi^\prime} 
    &\triangleq 
    b_{k} (\tilde{\tau}_{g_\tau}) b_k^\ast (\tilde{\tau}_{g_\tau^\prime}) 
    \mathbf{a}_{\mathrm{t}} (\tilde{\phi}_{g_\phi^\prime})
    \mathbf{a}_{\mathrm{t}}^\mathrm{H} (\tilde{\phi}_{g_\phi})
    c^\mathbf{\Omega}_{g_\tau, g_\tau^\prime, g_\phi, g_\phi^\prime} (\bar{\mathbf{X}}).
\end{align}

In this paper, the optimization problem in \eqref{eq:opt_unconst} is solved via \ac{Adam}~\cite{2014kingma_Adam}, with the derived gradients in \eqref{eq:grad_obj}, \eqref{eq:grad_sp}. 
The pseudocode of the proposed algorithm is provided in Algorithm~\ref{alg:prop}.
In Algorithm \ref{alg:prop}, $\eta$ and $T_\mathrm{iter}$ are the learning rate and the number of iterations of gradient descent, respectively. 
$\varepsilon$ is a hyper-parameter that prevents the objective variables from diverging for numerical stability. 
$\beta_1$ and $\beta_2$ are hyper-parameters that control the decay rates of the moving average of the gradient and the squared gradient, respectively. 
These hyper-parameters are set to $\beta_1=0.9,\ \beta_2=0.999,\ \varepsilon=10^{-8}$, which are commonly used as default settings~\cite{2014kingma_Adam}.

Based on the derived matrix $\bar{\mathbf{X}}$ via Adam algorithm, the pilot matrix $\mathbf{X}$, which satisfies the power constraint, is generated as $\mathbf{X} = \frac{\sqrt{P_\mathrm{t}}}{\| \bar{\mathbf{X}} \|_\mathrm{F}} \bar{\mathbf{X}}$ from \eqref{eq:X_bar}.
Since the derived pilot matrix $\mathbf{X}$ is a sparse matrix owing to the block sparse penalty in \eqref{eq:penalty_sp}, the subcarrier index set is obtained by
\begin{align}
    \label{eq:Q_opt}
    \mathcal{Q} = \left \{ k \ \big | \ \| \mathbf{X}_k \|_\mathrm{F}  \neq 0,\ \forall k \in \mathcal{K} \right \}.
\end{align}

Designing the sensing matrix $\mathbf{\Psi}$ with the derived pilot matrix $\mathbf{X}$ and subcarrier index set $\mathcal{Q}$, the channel estimation is performed.
The evaluation of the channel estimation performance with the optimized pilot design is described in Section~\ref{sec:simulation}.

%%%%%%%%%%%%%%%%%%%%%%%%
% Algorithm
% alg. code
\begin{algorithm}[t]
    \caption[]{Joint pilot allocation and sequence design}
    \label{alg:prop}
    \hrulefill
    \begin{algorithmic}[1]
        \vspace{-0.5ex}
        \Statex \textbf{Input:}  $\bar{\lambda},\ T_\mathrm{iter},\ \beta_1,\ \beta_2,\ \eta,\ \varepsilon$
        \Statex \textbf{Output:} $\mathcal{Q}=\{ k_1, \ldots, k_Q\}$, $\{ \mathbf{X}_k \}_{k_q \in \mathcal{Q}}$
        \vspace{-1.5ex}
        \Statex \hspace{-3ex} \hrulefill
        % \hrulefill

        %%%%%%%%%%%%%%%%%%%%%%%%%
        % % Initialize
        %%%%%%%%%%%%%%%%%%%%%%%%%
        \Statex \textbf{// Initialization} 
        \State 
            $\mathbf{M}_k^{(0)} = \mathbf{0}_{N_\mathrm{t} \times M}$, $\mathbf{V}_k^{(0)} = \mathbf{0}_{N_\mathrm{t} \times M},\ \forall k \in \mathcal{K}$, (First and second moments)
        \State 
            Generate initial pilot matrix $\bar{\mathbf{X}}^{(0)} = \begin{bmatrix} \bar{\mathbf{X}}_1^{(0)}, \ldots, \bar{\mathbf{X}}_K^{(0)} \end{bmatrix}$
        \State Calculate dictionaries $\mathbf{A}_\mathrm{t}(\tilde{\bm{\phi}})$, $\mathbf{B} (\tilde{\bm{\tau}})$ from \eqref{eq:At_dict}, \eqref{eq:B_dict}

        \Statex \textbf{// Gradient descent via Adam} 
        \For{$t=1, 2, \ldots, T_\mathrm{iter}$} $\ (\forall k \in \mathcal{K})$
            \State Calculate gradient $\mathbf{G}_k^{(t)} = \frac{\partial L(\bar{\mathbf{X}})}{\partial \bar{\mathbf{X}}_k^\ast}$ from \eqref{eq:grad_L}
            \Statex \hspace{0.4cm} // Update first and second moments
            \State $\mathbf{M}_k^{(t)} = \beta_1 \mathbf{M}_k^{(t-1)} + (1 - \beta_1) \mathbf{G}_k^{(t)}$
            \State $\mathbf{V}_k^{(t)} = \beta_2 \mathbf{V}_k^{(t)} + (1 - \beta_2)  ( \mathbf{G}_k^{(t) \ast} \mathbf{G}_k^{(t)}) $
            \State $\hat{\mathbf{M}}_k^{(t)} = \frac{1}{1 - (\beta_1)^t} \mathbf{M}_k^{(t)}$
            \State $\hat{\mathbf{V}}_k^{(t)} = \frac{1}{1 - (\beta_2)^t} \mathbf{V}_k^{(t)}$
            \Statex \hspace{0.4cm} // Update normalized pilot matrix
            \State 
            $\bar{\mathbf{X}}_k^{(t)} = \bar{\mathbf{X}}_k^{(t-1)} - \eta \left \{ \hat{\mathbf{M}}_k^{(t)} \oslash \left ( \sqrt{\hat{\mathbf{V}}_k^{(t)}} + \varepsilon \mathbf{1}_{N_\mathrm{t} \times M} \right ) \right \}$
        \EndFor
        \Statex \textbf{// Generate pilot matrices} 
        \State Calculate pilot matrix
        $\mathbf{X}_k = \frac{\sqrt{P_\mathrm{t}}}{\| \bar{\mathbf{X}} \|_\mathrm{F}} \bar{\mathbf{X}}_k, \ \forall k \in \mathcal{K}$
        \State Calculate subcarrier set 
        $\mathcal{Q} = \left \{ k \ \big | \ \| \mathbf{X}_k \|_\mathrm{F}  \neq 0,\ \forall k \in \mathcal{K} \right \}$
    \end{algorithmic}
\end{algorithm}

%%%%%%%%%%%%%%%%%%%%%%%%

\subsection{Complexity Evaluation}
\label{subsec:complexity}
The computational complexity of the objective functions and their gradients, evaluated in terms of the number of multiplications as \ac{FLOPs}, is shown in Table~\ref{table:FLOPs}.
Since the computation of $f^\mathbf{\Psi}(\mathbf{X})$ in \eqref{eq:fx} involves AoD, AoA, and delay, the complexity order increases quadratically with the product of the number of grids $G = G_\tau G_\phi G_\theta$.
In contrast, $f^\mathbf{\Omega}(\mathbf{X})$ can be calculated without requiring the dictionary for AoA, as shown in \eqref{eq:fx_omega}, therefore the complexity scales quadratically with the product of the number of grids $G_\tau G_\phi$, which is significantly lower than that of $f^\mathbf{\Psi}(\mathbf{X})$.
Accordingly, the gradient $\frac{\partial}{\partial \bar{\mathbf{X}}_k^\ast}
\left (\frac{f^\mathbf{\Omega}(\bar{\mathbf{X}})}{ \| \bar{\mathbf{X}}\|_\mathrm{F}^2} \right )$ can be efficiently computed independently of the AoA grids. 

% Table for complexity
\begin{table}[t!]
    \caption{Computational complexity} 
    \label{table:FLOPs}
    \centering
    \begin{tabular*}{8.5cm}{l|l}
        \hline
        \multicolumn{1}{c}{Functions} & FLOPs	\\
        \hline
        % Obj_psi
        $f^\mathbf{\Psi} (\mathbf{X})$
        & $\mathcal{O} \left ( N_\mathrm{r} M K G_\tau^2 G_\theta^2 G_\phi^2 + N_\mathrm{r}^2 N_\mathrm{t} M K^2 G_\tau G_\theta G_\phi \right )$ \\
        % Obj_omega
        $f^\mathbf{\Omega} (\mathbf{X})$
        & $\mathcal{O} \left ( N_\mathrm{t} M K G_\tau^2 G_\phi^2 \right )$ \\
        % g
        $g (\mathbf{X})$
        & $\mathcal{O} \left ( N_\mathrm{t} M K \right )$ \\
        % Grad_f_omega
        $\frac{\partial}{\partial \bar{\mathbf{X}}_k^\ast}
        \left (\frac{f^\mathbf{\Omega}(\bar{\mathbf{X}})}{ \| \bar{\mathbf{X}}\|_\mathrm{F}^2} \right )$
        & $\mathcal{O} \left (( M K  + N_\mathrm{t}) N_\mathrm{t}G_\tau^2 G_\phi^2 + N_\mathrm{t}^2 M K \right )$ \\
        % Grad_g
        $\frac{\partial}{\partial \bar{\mathbf{X}}_k^\ast} \left ( \frac{g(\bar{\mathbf{X}})}{\|\bar{\mathbf{X}}\|_\mathrm{F}} \right )$
        & $\mathcal{O} \left ( N_\mathrm{t} M K \right )$ \\
        \hline
    \end{tabular*}
\end{table}

\section{Numerical Results}
\label{sec:simulation}

\subsection{Simulation Setup}

% System param
This section evaluates the performance of the proposed pilot design under the following simulation parameters.
The carrier frequency is $f_\mathrm{c}=3.5\ \mathrm{GHz}$.
The system bandwidth and the number of subcarriers are $B = 1.92\ \mathrm{MHz}$\footnote{
For the sake of the compromise between simulation speed and performance validation, we chose this narrow band setup (i.e., approximately 5 \acp{PRB}) in this section. Notice that the generated sequences and allocation can be applied to different parts of the bandwidth or time symbols, such that one of the time or frequency or both dimensions of the generated sequence are repeated in other subbands or time symbols or both.
}
and $K=64$, respectively.
%%%%%%%%%%
% 
% 
The length of the pilot sequence is $M=8$.
The number of BS and UE antennas are $N_\mathrm{t}=32$ and $N_\mathrm{r}=8$, respectively.
The antenna spacing at the BS and UE sides are $d_\mathrm{t}=d_\mathrm{r}=\frac{\lambda_\mathrm{c}}{2}$.

% Channel param
The MIMO-OFDM channel, defined in \eqref{eq:H_k}, is generated by the composition of total $L=6$ paths, where
AoA $\phi_l$ and AoD $\theta_l$ are uniformly distributed in $[-\pi/2, \pi/2)$, and 
delay $\tau_l$ is uniformly distributed in $\left [0, (N_\mathrm{tap}-1)/B \right]$, as in~\cite{2017Venugopal_CE_delay_hybrid}, where $N_\mathrm{tap}=16$ is the number of delay taps.
The path gains are generated as
$\alpha_{l} \sim \mathcal{CN}(0, K_\mathrm{f} / (K_\mathrm{f} + 1)),\ (l=1,\ \mathrm{LoS})$ and
$\alpha_{l} \sim \mathcal{CN}(0, 1 / ((K_\mathrm{f} + 1)(L-1))),\ (l \neq 1,\ \mathrm{NLoS})$ with a Rician $K$-factor $K_\mathrm{f}=10\ \mathrm{dB}$~\cite{2016Gao_CE, 2018Lin_sparse_CE}.

% Grid design
The dictionary matrices for AoA, AoD, and delay in \eqref{eq:Ar_dict}-\eqref{eq:B_dict} are generated with the grids $\tilde{\theta}_{g_\theta}$, $\tilde{\phi}_{g_\phi}$, and $\tilde{\tau}_{g_\tau}$, which are computed as 
\begin{align}
    % theta
    \tilde{\theta}_{g_\theta} &= \sin^{-1} \left ( -1 + \frac{2}{G_\theta} (g_\theta - 1) \right ), \ g_\theta \in \{1, \ldots, G_\theta \}, \nonumber \\
    % phi
    \tilde{\phi}_{g_\phi} &= \sin^{-1} \left ( -1 + \frac{2}{G_\phi} (g_\phi - 1) \right ), \ g_\phi \in \{1, \ldots, G_\phi \}, \nonumber \\
    % 
    % tau
    \tilde{\tau}_{g_\tau} &=  \frac{(N_\mathrm{tap}-1) / B}{(G_\tau-1)} (g_\tau - 1) , \ g_\tau \in \{1, \ldots, G_\tau \}, \nonumber
\end{align}
where the number of grids are set to $G_\theta = 2 N_\mathrm{r}$, $G_\phi = 2 N_\mathrm{t}$, and $G_\tau = 2 N_\mathrm{tap}$ as in~\cite{2017Venugopal_CE_delay_hybrid}.

The parameters $p$ and $q$ for the generalized coherence in \eqref{eq:fx_omega} and the block-sparse penalty in \eqref{eq:penalty_sp} are empirically set to $p=4$ and $q=1$, respectively, since excessively large $p$ or excessively small $q$ may compromise the stability of the optimization process in gradient descent.
The learning rate $\eta$ and the number of iterations $T_\mathrm{iter}$ in the gradient descent algorithm are set to $\eta=10^{-3}$ and $T_\mathrm{iter}$ = 20,000.
The hyperparameter $\bar{\lambda}$ in \eqref{eq:opt_unconst} that controls the sparsity level in the pilot sequences is set in the range $[0.7,\ 7]$.
The initial pilot matrix $\bar{\mathbf{X}}^{(0)}$ in line 2 in Algorithm~\ref{alg:prop} is generated from i.i.d. complex Gaussian distribution as $\mathrm{vec}(\bar{\mathbf{X}}^{(0)}) \sim \mathcal{CN}(\mathbf{0}, \mathbf{I}_{N_\mathrm{t} M K})$.

% Metric
The channel estimation performance is evaluated by \ac{NMSE} under various signal-to-noise ratios (SNR).
SNR and NMSE are defined as follows:
\begin{align}
     & \mathrm{SNR} \triangleq \frac{P_\mathrm{t} / (N_\mathrm{t} M Q)}{\sigma^2},
\end{align}
\begin{align}
    & \mathrm{NMSE}(\bm{\mathbf{h}}) \triangleq \mathbb{E} \left[ {\| \bm{\mathbf{h}} - \hat{\bm{\mathbf{h}}} \|_2^2} / {\| \bm{\mathbf{h}} \|_2^2} \right],
\end{align}
where $\mathbf{h}$ and $\hat{\mathbf{h}}$ are the true channel vector in \eqref{eq:h_vec} and the estimated channel vector in \eqref{eq:h_vec_est}, respectively. 
To evaluate the impact of pilot design on channel estimation performance, CS-based channel estimation methods OMP~\cite{1993Pati_OMP} and \ac{GAMP-SBL}~\cite{2018Shoukairi_GAMP_SBL} are used.

% Methods
To evaluate the effectiveness of the proposed pilot design, the following methods are compared in the simulation.
\begin{itemize}
    \item[1)] \textbf{\textit{Gauss + Rand.}} : 
    The pilot sequence is generated from i.i.d. complex Gaussian distribution, and the subcarrier allocation $\mathcal{Q}$ is randomly determined. 
    As demonstrated in~\cite{2005Candes_CS_GaussRand}, randomly generated sensing matrices guarantee exact sparse recovery with high probability based on the RIP.
    Thus, the use of randomly generated pilots is statistically optimal.
    Note that the implementation of such pilots in practical systems is challenging.
    This approach is a baseline pilot design for comparison.
    \item[2)] \textbf{\textit{SIDCO + GA}} : 
    The subcarrier allocation $\mathcal{Q}$ is optimized by GA~\cite{2016He_allocation_MIMO_OFDM}, and the pilot sequence is optimized by SIDCO~\cite{2022Iimori_bilinear_Grant_free} given the subcarrier allocation $\mathcal{Q}$.
    This approach is a combination of the conventional sequence design and subcarrier allocation. 
    The sequence design~\cite{2022Iimori_bilinear_Grant_free} assumes that all subcarriers share the same SIDCO sequence. 
    To reduce coherence in the frequency domain, this method is extended as follows.
    First, we construct the SIDCO sequence $\mathbf{X}_{k_0}$ for the $k_0$-th subcarrier, and the sequences for the remaining subcarriers $ \{ \mathbf{X}_k \}_{\forall k \in \mathcal{Q} \setminus k_0}$, are then generated by randomly permuting the row vectors of $\mathbf{X}_{k_0}$.
    \item[3)] \textbf{\textit{Opt. sequence + GA}} : 
    The subcarrier allocation is optimized by GA~\cite{2016He_allocation_MIMO_OFDM}, and the pilot sequence is optimized by the proposed optimization given the subcarrier allocation $\mathcal{Q}$ without using the block sparse penalty $g(\mathbf{X})$ in \eqref{eq:penalty_sp}. 
    This approach is a combination of the conventional subcarrier allocation and the proposed sequence design.
    \item[4)] \textbf{\textit{Prop. (Opt. sequence + Opt. alloc.)}} : 
    The pilot sequence and subcarrier allocation are jointly optimized by the proposed method. 
\end{itemize}

In what follows, we evaluate the optimized pilot sequence in Section~\ref{subsec:eval_pilot} and evaluate the performance of channel estimation using the designed pilot in Section~\ref{subsec:eval_CE}.

\subsection{Evaluation of the Optimized Pilot}
\label{subsec:eval_pilot}

%%%%%%%%%%%%%%%%%%%%%%%%%%%
% Color map
\begin{figure}[t]
    % Gauss
    \begin{minipage}{1.0\columnwidth}
        \centering
        \includegraphics[width=\linewidth]{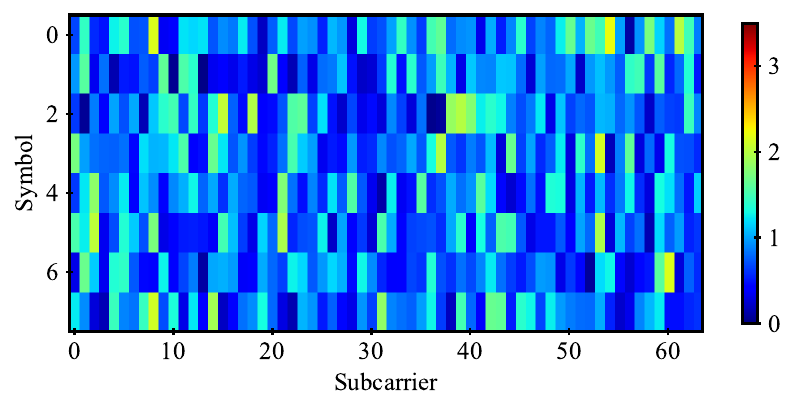}
        \vspace{-4ex}
        \subcaption{Initial pilot sequence.} 
        \label{fig:Cmap_Gauss}
    \end{minipage} 
    \\
    % 
    % Prop.
    \begin{minipage}{1.0\columnwidth}
        \centering
        \includegraphics[width=\linewidth]{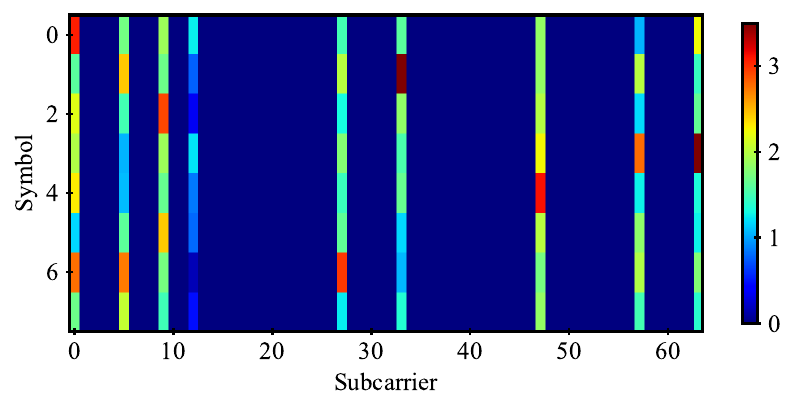}
        \vspace{-4ex}
        \subcaption{Optimized pilot sequence.} 
        \label{fig:Cmap_Opt}
    \end{minipage}
    \caption{
    The amplitude of the pilot sequences at the first transmit antenna in the time-frequency domain: (a) Initial pilot sequence, generated from i.i.d. complex Gaussian distribution. (b) Optimized pilot sequence, obtained by Algorithm~\ref{alg:prop}, where the hyper-parameter is set to $\bar{\lambda}=1.5$, resulting in the number of pilot subcarriers being $Q=9$.
    }
    \label{fig:Cmap}
\end{figure}

To visualize the designed sequence and allocation, Figs.~\ref{fig:Cmap_Gauss} and \ref{fig:Cmap_Opt} depict the amplitude of the initial and optimized pilot sequences at the first transmit antenna in the time-frequency domain.
Since the initial pilot sequence over all $K$ subcarriers is generated from i.i.d. complex Gaussian distribution, the pilot matrix $\mathbf{X}^{(0)}$ is a dense matrix, as shown in Fig.~\ref{fig:Cmap_Gauss}. 
Using this initial pilot as a starting value of the gradient descent algorithm, the optimized pilot matrix $\mathbf{X}$ becomes a sparse matrix owing to the block sparse penalty in \eqref{eq:penalty_sp}, as shown in Fig.~\ref{fig:Cmap_Opt}.
Since the sparse penalty accounts for block sparsity in the antenna direction, each transmit antenna possesses the same sparse structure in Fig.~\ref{fig:Cmap_Opt}.
As illustrated in Fig.~\ref{fig:Cmap_Opt}, the number of pilot subcarriers excluding zero elements becomes $Q=9$ when the hyper-parameter is set to $\bar{\lambda}=1.5$.
% 
%%%%%%%%%%%%%%%%%%%%%%%%%%%

%%%%%%%%%%%%%%%%%%%%%%%%%%%
% Loss, Q_vs_lambda
\begin{figure}[t]
    % Loss_vs_iterations
    \begin{minipage}{1.0\columnwidth}
        \centering
        \includegraphics[width=\linewidth]{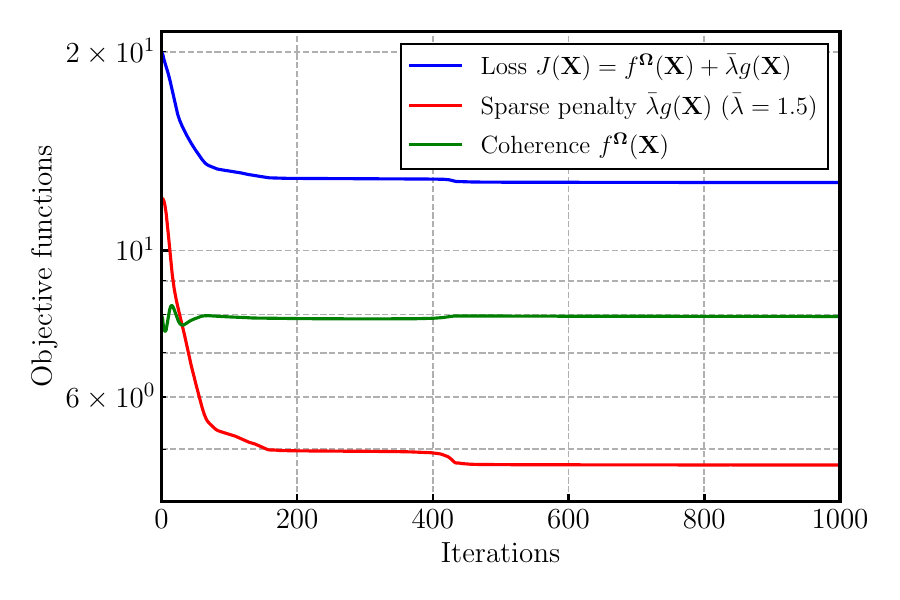}
        \vspace{-5ex}
        \caption{Objective functions, $J(\mathbf{X})$, $\bar{\lambda} g(\mathbf{X})$, and $f^\mathbf{\Omega} (\mathbf{X})$, against algorithmic iterations. The hyper-parameter$\bar{\lambda}$ is set to 1.5, which corresponds to $Q=9$.} 
        \label{fig:Loss_vs_iterations}
    \end{minipage}
    % 
    % Q_vs_lambda
    \begin{minipage}{1.0\columnwidth}
        \centering
        \includegraphics[width=\linewidth]{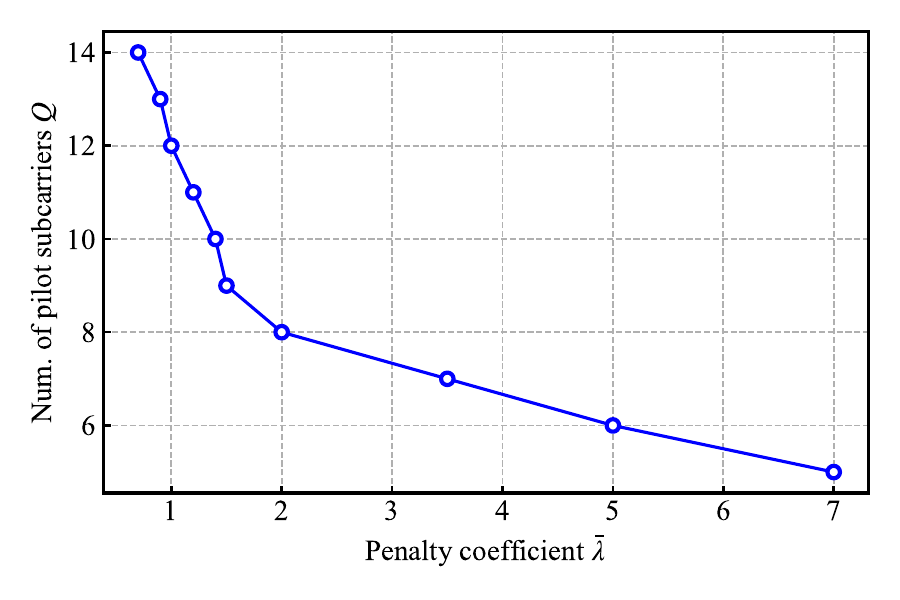}
        \vspace{-5ex}
        \caption{The number of pilot subcarriers $Q$ versus penalty coefficient $\bar{\lambda}$.} 
        \label{fig:Q_vs_lambda}
    \end{minipage}
\end{figure}

Fig.~\ref{fig:Loss_vs_iterations} shows the objective functions, $L(\mathbf{X})$, $\bar{\lambda} g(\mathbf{X})$, and $f^\mathbf{\Omega}(\mathbf{X})$, against algorithmic iterations in the optimization algorithm.
The loss function $L(\mathbf{X})$, composed of the summation of the coherence metric $f^\mathbf{\Omega}(\mathbf{X})$ and the block sparse penalty $\bar{\lambda} g(\mathbf{X})$, decreases with iterations via the gradient-descent approach.
Although coherence generally rises as sparsity increases due to a reduction in the degrees of freedom of the pilot matrix, the proposed method improves sparsity level while minimizing the increase in coherence.
The sparsity level, corresponding to the number of pilot subcarriers $Q$, depends on the hyper-parameter $\bar{\lambda}$.

% Q_vs_lambda
To evaluate the relationship between the hyperparameter $\bar{\lambda}$ and the number of pilot subcarriers $Q$, Fig.~\ref{fig:Q_vs_lambda} shows the number of subcarriers $Q$ as a function of $\bar{\lambda}$.
As shown in the figure, the number of subcarriers $Q$ decreases as the hyperparameter $\bar{\lambda}$ increases, resulting in a reduction in pilot overhead at the expense of channel estimation performance. 
The evaluation of channel estimation performance is described in the following subsection.
%%%%%%%%%%%%%%%%%%%%%%%%%%%

%%%%%%%%%%%%%%%%%%%%%%%%%%%
% CDF
\begin{figure}[t]
    % Coherence
    \begin{minipage}{1.0\columnwidth}
        \centering
        \includegraphics[width=\linewidth]{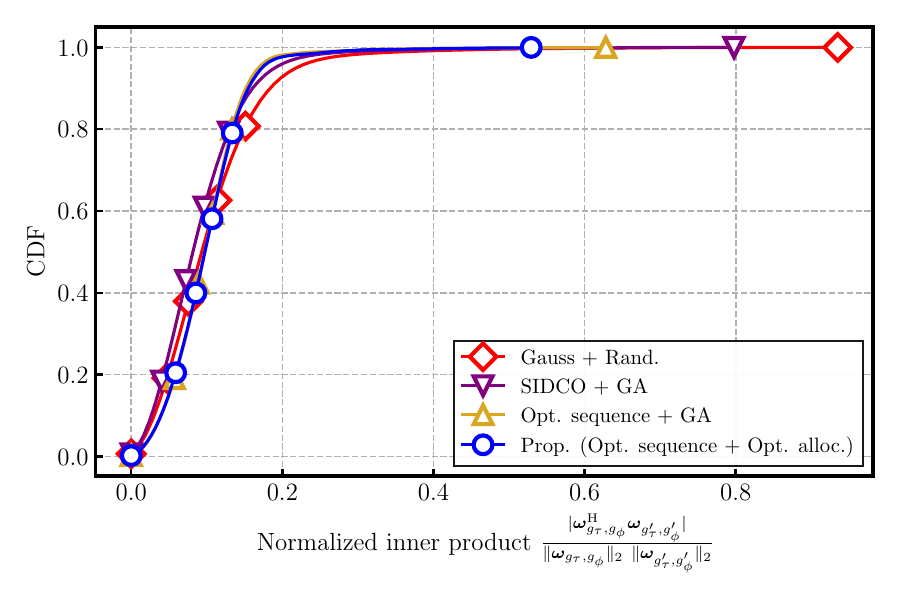}
        \vspace{-5ex}
        \caption{The CDF of the normalized inner products between the column vectors of $\mathbf{\Omega}$ when $Q=9$.} 
        \label{fig:CDF_of_coherence_Q9}
    \end{minipage}
    % 
    % Norm
    \begin{minipage}{1.0\columnwidth}
        \centering
        \includegraphics[width=\linewidth]{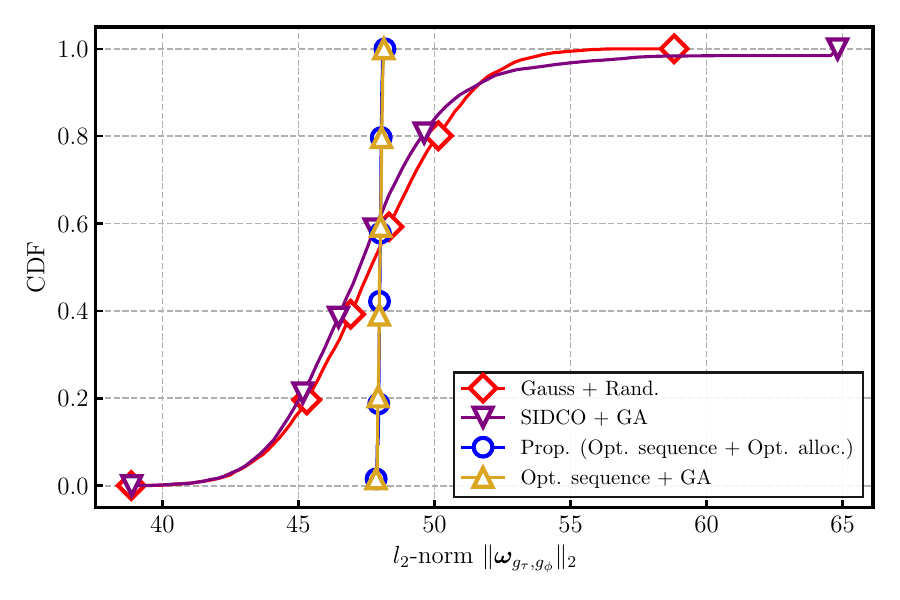}
        \vspace{-5ex}
        \caption{The CDF of the $\ell_2$-norm $\| \bm{\omega}_{g_\tau, g_\phi}\|_2$ when $Q=9$.} 
        \label{fig:CDF_of_norm_Q9}
    \end{minipage}
\end{figure}

% CDF_of_coherence
For the evaluation of the coherence performance of the sensing matrix $\mathbf{\Psi}$ given the designed pilot matrix $\mathbf{X}$, we offer, in Fig.~\ref{fig:CDF_of_coherence_Q9}, the \ac{CDF} of the inner products of the column vectors of $\mathbf{\Omega}$.
As explained in \eqref{eq:c_psi_omega}, since the coherence of the sensing matrix $\mathbf{\Psi}$ is determined by the coherence of $\mathbf{\Omega}$, the normalized inner product $\frac{ |\boldsymbol{\omega}_{g_\tau, g_\phi}^\mathrm{H} \boldsymbol{\omega}_{g_\tau^\prime, g_\phi^\prime}| } {\|\boldsymbol{\omega}_{g_\tau, g_\phi}\|_2 \ \|\boldsymbol{\omega}_{g_\tau^\prime, g_\phi^\prime}\|_2}$ is evaluated in Fig~\ref{fig:CDF_of_coherence_Q9}.
As shown in the figure, the proposed method reduces high-value inner products more effectively compared to other pilot designs.
Compared to \textit{Opt. Sequence + GA}, which designs subcarrier allocation and pilot sequence independently, the proposed method, which jointly optimizes both, can reduce the largest inner product, corresponding to the mutual coherence.

% CDF_of_norm
Fig.~\ref{fig:CDF_of_norm_Q9} shows the \ac{CDF} of the $\ell_2$-norm of the column vector of $\mathbf{\Omega}$.
Since the $\ell_2$-norm of the column vector of $\mathbf{\Psi}$ can be expressed, from \eqref{eq:c_psi_omega}, as 
$ \| \bm{\psi}_g \|_2 = \| \bm{\omega}_{g_\tau, g_\phi} \|_2 \cdot \| \mathbf{a}_\mathrm{r} (\tilde{\theta}_{g_\theta})\|_2 = N_\mathrm{r} \| \bm{\omega}_{g_\tau, g_\phi} \|_2$, 
we evaluate the norm $\| \bm{\omega}_{g_\tau, g_\phi} \|_2$ in Fig.~\ref{fig:CDF_of_norm_Q9}.
As shown in the figure, the $\ell_2$-norm in the proposed sequence design approaches an almost identical value, such that $\| \bm{\psi}_1 \|_2 \simeq \cdots \simeq \| \bm{\psi}_G \|_2$. 
Thus, the minimization of the transformed generalized coherence $f^\mathbf{\Psi} (\mathbf{X})$ in \eqref{eq:fx} leads to the minimization of the generalized coherence $\nu_p(\mathbf{\Psi})$ in \eqref{eq:GC}, thereby improving the CDF of the normalized inner product of $\mathbf{\Omega}$, as shown in Fig.~\ref{fig:CDF_of_coherence_Q9}.
%
%%%%%%%%%%%%%%%%%%%%%%%%%%%

\subsection{Evaluation of Channel Estimation Performance}
\label{subsec:eval_CE}

%%%%%%%%%%%%%%%%%%%%%%%%%%%
% NMSE_vs_SNR
\begin{figure}[t]
    % OMP
    \begin{minipage}{1.0\columnwidth}
        \centering
        \includegraphics[width=\linewidth]{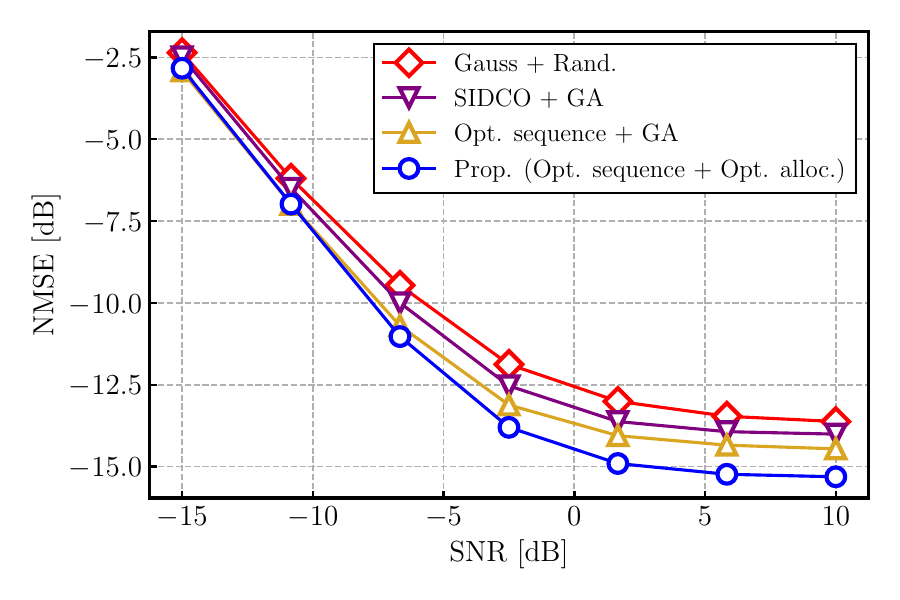}
        \vspace{-4ex}
        \subcaption{OMP}
        \label{fig:NMSE_vs_SNR_OMP_Q9}
    \end{minipage} 
    % 
    % SBL
    \begin{minipage}{1.0\columnwidth}
        \centering
        \includegraphics[width=\linewidth]{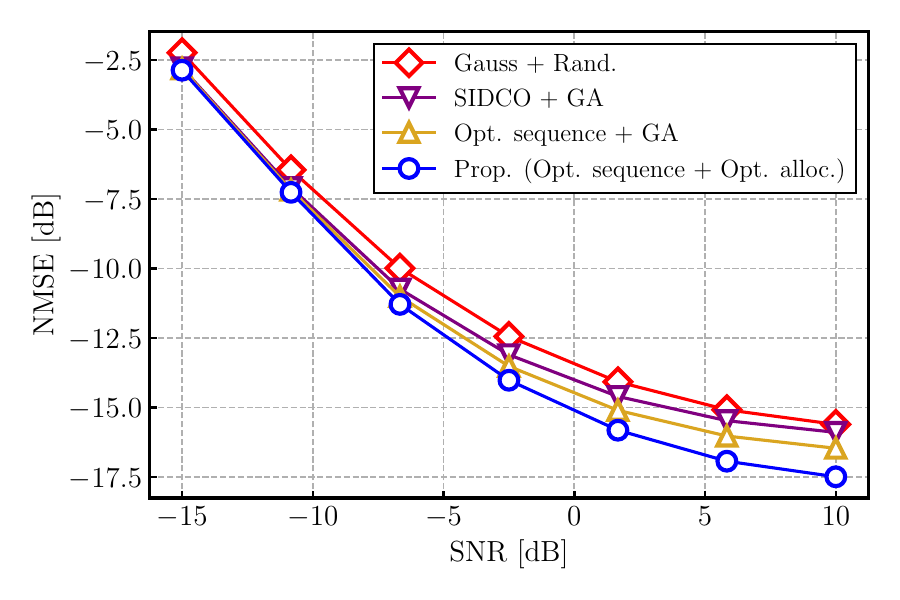}
        \vspace{-4ex}
        \subcaption{GAMP-SBL} 
        \label{fig:NMSE_vs_SNR_SBL_Q9}
    \end{minipage}
    \caption{NMSE versus SNR when $Q=9$. Channel estimation is performed by compressed sensing algorithms, (a) OMP and (b) GAMP-SBL.}
    \label{fig:NMSE_vs_SNR_Q9}
\end{figure}

Fig.~\ref{fig:NMSE_vs_SNR_Q9} shows the NMSE as a function of SNR, where (a) and (b) illustrate the NMSE when the OMP~\cite{1993Pati_OMP} and GAMP-SBL~\cite{2018Shoukairi_GAMP_SBL} algorithms are employed as channel estimation methods.  
As shown in these figures, the combination of the conventional methods, \textit{SIDCO + GA}, outperforms the naive method, \textit{Gauss + Rand.}, which indicates that designing the pilot sequence and subcarrier allocation accounting for the coherence metric can enhance channel estimation performance when using CS-based approaches.
Comparing \textit{SIDCO + GA} and \textit{Opt. sequence + GA}, it can be seen that the proposed sequence design, given the subcarrier allocation determined by GA, further enhances channel estimation performance.
This improvement arises from the fact that the proposed sequence design enhances coherence performance by reducing high-value inner products, as shown in Fig.~\ref{fig:CDF_of_coherence_Q9}.
However, the performance of \textit{Opt. sequence + GA}  is limited due to the separate design of subcarrier allocation and pilot sequence.
In contrast, the proposed method, \textit{Prop. (Opt. sequence + Opt. alloc.)}, jointly optimizes subcarrier allocation and pilot sequence, leading to further improvements in channel estimation performance.
As illustrated in Fig.~\ref{fig:NMSE_vs_SNR_OMP_Q9} and Fig.~\ref{fig:NMSE_vs_SNR_SBL_Q9}, the proposed method, \textit{Prop. (Opt. sequence + Opt. alloc.)}, consistently outperforms other pilot designs across all SNR region and under both CS methods, OMP and GAMP-SBL.
% 
%%%%%%%%%%%%%%%%%%%%%%%%%%%

%%%%%%%%%%%%%%%%%%%%%%%%%%%
% NMSE_vs_Q
\begin{figure}[t]
    % OMP
    \begin{minipage}{1.0\columnwidth}
        \centering
        \includegraphics[width=\linewidth]{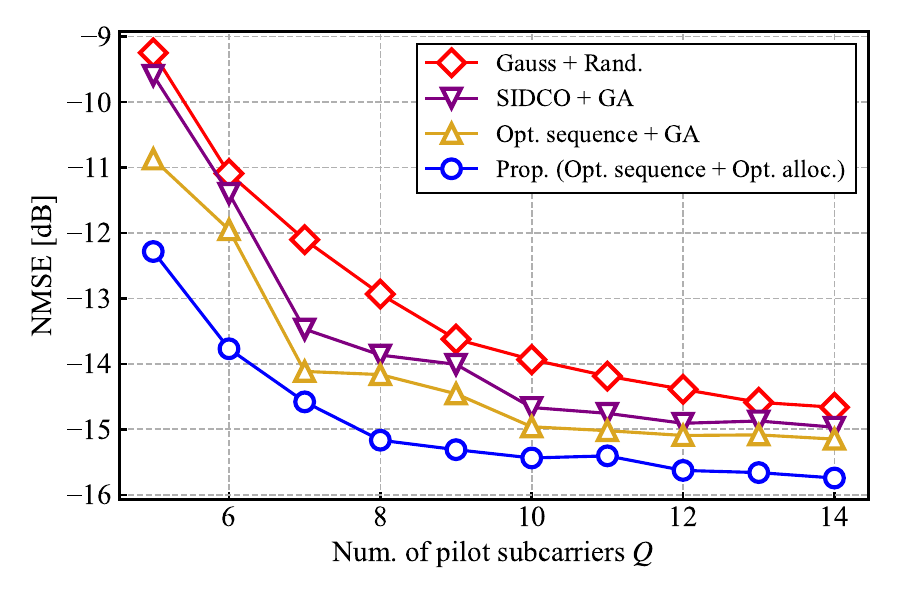}
        \vspace{-4ex}
        \subcaption{OMP} 
        \label{fig:NMSE_vs_Q_OMP}
    \end{minipage} 
    % 
    % SBL
    \begin{minipage}{1.0\columnwidth}
        \centering
        \includegraphics[width=\linewidth]{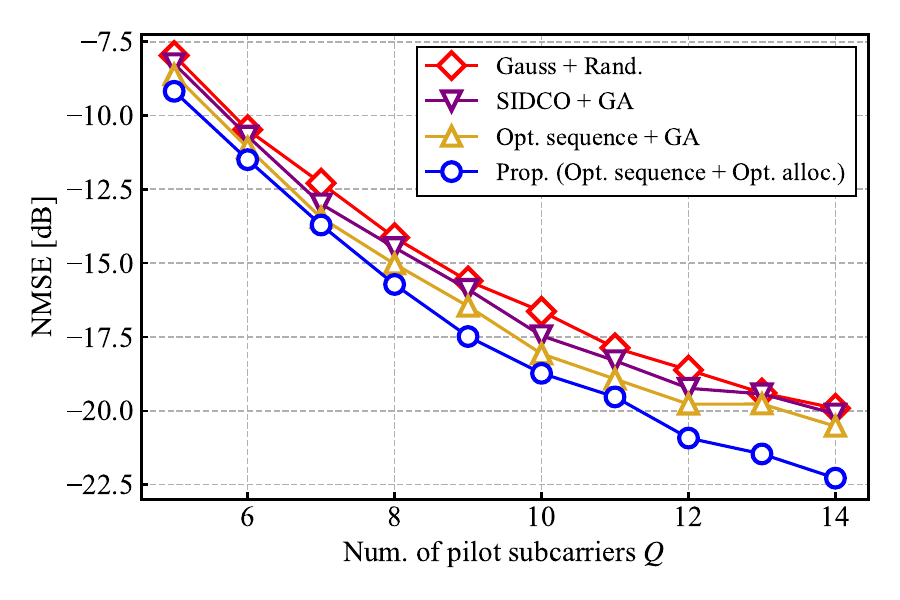}
        \vspace{-4ex}
        \subcaption{GAMP-SBL} 
        \label{fig:NMSE_vs_Q_SBL}
    \end{minipage}
    \caption{NMSE versus the number of pilot subcarriers $Q$ when $\mathrm{SNR}=10\ \mathrm{dB}$. Channel estimation is performed by compressed sensing algorithms, (a) OMP and (b) GAMP-SBL}
    \label{fig:NMSE_vs_Q}
\end{figure}

Fig.~\ref{fig:NMSE_vs_Q} presents the NMSE versus the number of pilot subcarriers $Q$ at $\mathrm{SNR}=10 \ \mathrm{dB}$.
The number of pilot subcarriers $Q$ is varied by controlling the hyperparameter $\bar{\lambda}$, as shown in Fig.~\ref{fig:Q_vs_lambda}.
As illustrated in the figure, the naive pilot design, \textit{Gauss + Rand.}, exhibits the worst performance for all $Q$, indicating that optimization of subcarrier allocation and pilot sequence can enhance channel estimation performance.
A comparison of \textit{SIDCO + GA} and \textit{Opt. sequence + GA} demonstrates the performance gain achieved through the sequence design via the proposed method.
Among the pilot designs, the proposed method, \textit{Prop. (Opt. sequence + Opt. alloc.)}, achieves the best performance for any number of pilot subcarriers $Q$ under both CS approaches, OMP and GAMP-SBL. 
%%%%%%%%%%%%%%%%%%%%%%%%%%%

\section{Conclusion}
\label{sec:conclusion}
In this paper, we propose a joint optimization of pilot subcarrier allocation and non-orthogonal sequence design for MIMO-OFDM systems.
To improve the channel estimation accuracy in CS-based frameworks that exploit delay-angle sparsity, we formulate an optimization problem that minimizes the coherence metric of the sensing matrix.
To address the intractability of discrete variables in subcarrier allocation, we introduce a block sparse penalty to eliminate unnecessary pilot sequences in the frequency domain, which enables joint optimization using only continuous variables, without relying on discrete variables.
For improved computational efficiency, the coherence metric is reformulated as a compact expression that is independent of the AoA grids.
The gradient is then analytically derived in closed-form, facilitating the efficient solution of the joint optimization problem via a gradient descent approach.
Moreover, computer simulations reveal that the pilot designed by the proposed optimization exhibits superior coherence properties compared to existing pilot designs and improves CS-based channel estimation performance.

\bibliographystyle{IEEEtran}
\bibliography{reference.bib}

% % %%%%%%%%%%%%%%%%
% % % Biography
% % %%%%%%%%%%%%%%%%
% \input{fig/bio/bio_tex}
% % %%%%%%%%%%%%%%%%%

\end{document}